\documentclass[%
preprintnumbers,
nofootinbib,
 amsmath,amssymb,
 aps,
 prd,
 twocolumn,
 superscriptaddress,
]{revtex4-1}

\usepackage{dcolumn}

\allowdisplaybreaks

\usepackage[colorlinks=true, pdfstartview=FitV, linkcolor=red,citecolor=green, bookmarks=true, bookmarksnumbered=true,breaklinks]{hyperref}
\usepackage[dvipdfmx]{graphicx}
\usepackage{latexsym,amsfonts,amssymb,amsmath,bm,mathrsfs} 

\newcommand{\id}{\mbox{1}\hspace{-0.25em}\mbox{l}}

\newcommand{\tr}{ {\rm Tr} }
\newcommand{\Lag}{ {\mathscr{L}} }

\newcommand{\M}{ {\mathcal M} }

\newcommand{\ext}{ {\rm ext} }

\newcommand{\vac}{ {\rm vac} }

\newcommand{\B}{ {\mathcal B} }

\newcommand{\EM}{ {\rm em} }

\newcommand{\scJ}{ {\scriptstyle  J} }

\newcommand{\pv}{{_{\scriptscriptstyle  \rm PV}}}

\newcommand{\ps}{ {\scriptscriptstyle \rm P} }
\newcommand{\V}{ {\scriptscriptstyle \rm V} }

\newcommand{\gam}{ {\scriptscriptstyle \gamma} }

\newcommand{\mix}{ {\rm mix} }

\newcommand{\ml}{{ m_l }}
\newcommand{\mh}{{ m_h }}
\newcommand{\Ql}{{ Q_l }}
\newcommand{\Qh}{{ Q_h }}
\newcommand{\gtt}{\gamma^1\gamma^2}

\newcommand{\eB}{{(eB)^2}}

\newcommand{\Dstar}{ {D^\ast} }
\newcommand{\Dn}{ {\scriptscriptstyle D^0} }
\newcommand{\Dpm}{ {\scriptscriptstyle D^\pm} }

\newcommand{\beq}{\begin{eqnarray}}
\newcommand{\eeq}{\end{eqnarray}}

\allowdisplaybreaks[1]

\usepackage[normalem]{ulem}  

\newcommand{\changed}[1]{{\sf\color[rgb]{1,0,0}{#1}}}

\usepackage{slashed}

\begin{document}

\preprint{KEK-TH-1883, RBRC-1162, RIKEN-QHP-212} 

\vspace*{-10mm}
\begin{flushright}
\end{flushright}
\vspace{-5mm}


\author{Philipp Gubler} \email{\tt pgubler@riken.jp}

\affiliation{ECT$^*$, Villa Tambosi, 38123 Villazzano (Trento), Italy}


\author{Koichi Hattori} \email{\tt koichi.hattori@riken.jp}

\affiliation{
RIKEN BNL Research Center, Building 510A, Brookhaven National Laboratory, Upton, New York 11973, USA
}
\affiliation{
Theoretical Research Division, Nishina Center, RIKEN, Wako, Saitama 351-0198, Japan
}


\author{Su Houng Lee} \email{\tt suhoung@yonsei.ac.kr}

\affiliation{
Department of Physics and Institute of Physics and Applied Physics,
Yonsei University, Seoul 120-749, Korea
}


\author{Makoto Oka} \email{\tt oka@th.phys.titech.ac.jp}

\affiliation{
Department of Physics, Tokyo Institute of Technology, Meguro, Tokyo 152-8551, Japan
}
\affiliation{
Advanced Science Research Center, Japan Atomic Energy Agency, 
Tokai, Ibaraki, 319-1195 Japan
}


\author{Sho Ozaki} \email{\tt sho@post.kek.jp}

\affiliation{
Theory Center, IPNS, High Energy Accelerator Research Organization (KEK),
1-1 Oho, Tsukuba, Ibaraki 305-0801, Japan
}


\author{Kei Suzuki} \email{\tt kei.suzuki@riken.jp}

\affiliation{
Theoretical Research Division, Nishina Center, RIKEN, Wako, Saitama 351-0198, Japan
}


\vspace*{5mm}
\title{$D$ mesons in a magnetic field}


\date{\today}

\begin{abstract}
We investigate the mass spectra of open heavy flavor mesons 
in an external constant magnetic field within QCD sum rules. 
Spectral {\it Ans\"atze} on the phenomenological side are proposed 
in order to properly take into account mixing effects between the pseudoscalar and vector channels, 
and the Landau levels of charged mesons. 
The operator product expansion is implemented up to dimension-5 operators. 
As a result, we find for neutral $D$ mesons a significant positive mass shift that goes beyond simple 
mixing effects. 
In contrast, charged $D$ mesons are further subject to Landau level effects, 
which together with the mixing effects 
almost completely saturate the mass shifts obtained in our sum rule analysis. 
\end{abstract}

\maketitle
%
%

\section{Introduction}

Strong electromagnetic fields created in ultrarelativistic heavy-ion collisions \cite{Bestimates} 
and neutron stars/magnetars \cite{TD, HL06}
have motivated a number of phenomenological studies that lead to the discovery of  novel  
phenomena such as the chiral magnetic effect \cite{KMW} and the (inverse) magnetic catalysis \cite{GMS} 
(see Refs.~\cite{KLSY, Kha14, Tuc14, ANT, MS15} for recent reviews and references therein).
Recently, using an effective field theory method, the $\rho$-meson condensation
was also discussed \cite{Chern}. While the zero mode proposed to emerge in the $\rho$-meson channel
has not been supported by QCD-based arguments, lattice QCD simulations
or quark model calculations \cite{HY, rho_lat, taya, hks},
these studies have shown intriguing dynamics of the composite particles
which are quite different from the naive picture on the basis of the hadronic degrees of freedom,
and thus suggest the importance of taking into account the elementary degrees of freedom,
i.e., quarks and gluons, when the strength of the magnetic field becomes larger than the QCD scale.

The QCD sum rule (QCDSR) is a semianalytic method \cite{SVZ792, RRYrev, Shifman, Narison, CK2001} 
which allows one to investigate the hadron mass spectrum on the basis of QCD  
using the dispersion relation for a current correlator. 
With the help of the operator product expansion (OPE) \cite{Wil69}, one can include nonperturbative properties of the QCD vacuum, as well as 
perturbative dynamics of quarks and gluons, into 
a power series expansion of the  current correlator. 
The information encoded in the OPE is then connected to the spectral density 
on the other side of the dispersion relation called the phenomenological side. 
After having been developed for the  vacuum \cite{SVZ792, RRY80,RRY81}, QCDSR was applied to hadron properties at finite temperature and density \cite{BS86}. 
Some of the present authors investigated 
light vector mesons \cite{Furnstahl-Hatsuda-Lee90, HatsudaLee, PO14, GW15} 
and quarkonium spectra 
\cite{Furnstahl-Hatsuda-Lee90, KKLW, ML, GMO, SGMO, Lee:2013dca} 
in media which are related to important issues in heavy-ion collisions  
\cite{MS, Hashi}. 
Open heavy flavor systems have also been investigated in vacuum \cite{HL_vac} and finite density \cite{HL, ZHK, Suzuki_thesis}.

Recently, we have extended the framework of QCDSR, applying it to charmonium spectroscopy in magnetic fields \cite{Letter, PRD}, 
where the effects of magnetic fields were implemented both on the OPE side and the phenomenological side. 
Investigating the roles of magnetically induced terms on the phenomenological side, 
we have constructed a spectral {\it Ansatz} that takes into account the mixing effects 
between $\eta_c$ and $J/\psi$ induced by a magnetic field.  
By using this {\it Ansatz}, we have obtained mass spectra 
consistent with those from potential model calculations \cite{AS,Bonati:2015dka}. 
Therefore, it is now understood that one should use the correctly modified 
spectral {\it Ansatz} not only for the charmonia 
but also for general hadrons when the mixing effects emerge.

In this work, we investigate mass spectra of pseudoscalar open heavy flavors 
in the presence of external constant magnetic fields. 
Here again, we take into account the effects of magnetic fields 
both on the OPE side and on the phenomenological side. 
In contrast to charmonia, open heavy flavors include not only 
electrically neutral but also charged mesons, whose masses will split in magnetic fields. 
While the mixing effects occur to both of them, 
charged mesons are subject to another effect, namely 
the Landau level quantization in magnetic fields. 
In this paper, we will discuss how to properly treat the Landau levels on the phenomenological side for the first time. 
On the OPE side, we take into account operators up to dimension 5 for the vacuum part and 
compute the Wilson coefficients of operators up to dimension 4 
and second order in $eB$ for terms depending on the magnetic field. 
This expansion is valid when the magnitude of the magnetic field 
is small compared to the separation scale of the OPE\footnote{
More precisely, the magnitude of magnetic fields should be 
compared with the Borel mass since it works as a separation scale 
as discussed below (see Appendix~\ref{sec:pertapp}). }.  
New operator expectation values appear  in the presence of external magnetic fields as seen 
in the tensor-type quark condensate $\langle \bar q \sigma^{\mu\nu} q \rangle$ \cite{IS, sus_lat2}. 

We should  mention here a preceding work \cite{Machado} 
in which $B$-meson spectra in magnetic fields were investigated by using QCDSR. 
While we follow their basic strategy for the implementation of the OPE, 
we would like to point out two 
important differences between their and our results. 
First, in Ref.~\cite{Machado}, the mixing effects between $B$ and $B^\ast$ mesons 
were not taken into account on the phenomenological side. 
This mixing effect could be strong, 
when the mass difference between the mixing partners is small,  as 
is indeed  the case for the  $B$ and $B^\ast$ mesons. 
Second, the emergence and exact cancellation of infrared (IR) divergences on the OPE side 
were not treated properly in Ref.~\cite{Machado}. In the chiral limit, 
infrared divergences appear in quark loops with insertions of constant magnetic fields, 
due to soft (zero) momentum transfers from the external constant field. 
However, as shown in Ref.~\cite{ZHK} at finite density and in the present work for magnetic fields, 
the IR divergences between the quark loops and the quark condensates exactly cancel, leading to an IR-finite OPE. 
To achieve this, it is necessary to consistently include all types of quark condensates 
up to the second order in the magnetic field. We show that 
a dimension-4 condensate should be included as well as 
the dimension-3 condensates considered in Ref.~\cite{Machado}. 
We will come back to these points below.

This manuscript is organized as follows: We first briefly describe the conventional QCDSR analysis in Sec.~\ref{sec:QCDSR}, 
and then examine the phenomenological side of the $D$-meson channel in magnetic fields in Sec.~\ref{sec:ph} and the corresponding OPE side in Sec.~\ref{sec:OPE}. 
The mass spectra obtained from the QCDSR analysis are given in Sec.~\ref{sec:result}.
Section~\ref{sec:summary} is devoted to the summary of this work. 
Appendixes include estimates of the mixing strengths, 
 the treatment of the infrared divergences, 
estimates of the condensates, and tables of the Borel-transformed OPE.


\section{Brief description of QCD sum rules}

\label{sec:QCDSR}

We define a time-ordered current correlator by 
\begin{eqnarray}
\Pi^J(q) = i \int \!\! d^4x \, e^{iqx} \langle 0 \vert T[ J(x) J(0) ] \vert 0 \rangle
\label{eq:JJ}
\ ,
\end{eqnarray}
where the superscript $J$ specifies a channel. 
We investigate neutral and charged 
open heavy flavors composed of a light quark and a heavy antiquark 
and their antiparticles: 
\begin{eqnarray}
{\rm Neutral }: && \ J^{(0)} = i \bar u \gamma^5 c , \ J^{(\bar 0)} = i \bar c \gamma^5 u
\\
{\rm Charged }: && \ J^{(+)} = i \bar d \gamma^5 c , \ J^{(-)} = i \bar c \gamma^5 d
.
\end{eqnarray}
While we will focus on charmed mesons throughout this paper, 
extension to bottom mesons is straightforward. 
The mass spectra of the two neutral 
and separately the two charged mesons 
remain degenerate in an external magnetic field. 
The QCD sum rule relates the hadronic spectral function in the physical region ($q^2 = - Q^2 \geq 0$) 
to the operator product expansion (OPE) performed 
in the deep Euclidean region ($Q^2 \to \infty$), 
through the dispersion relation given as 
\begin{eqnarray}
\Pi^\scJ( Q^2 )  = \frac{1}{\pi} \int_0^\infty
\frac{ \, {\rm Im} \, \Pi^\scJ (s) \, }{ s + Q^2 } \, ds
\ + \ ({\rm subtractions})
\label{eq:disp0}
.
\end{eqnarray}
To extract the lowest-lying pole in the spectral function 
$\rho(s) = \pi^{-1} {\rm Im} \, \Pi^\scJ (s) $, 
one can use an {\it Ansatz} conventionally called ``pole + continuum'' given by 
\begin{eqnarray}
\rho_{\rm vac}(s) =
f_0 \delta (s -m ^2) +  \frac{1}{\pi} {\rm Im} \, \Pi^\scJ_{\rm pert} (s) \theta(s - s_{\rm th}). 
\label{eq:pole}
\end{eqnarray}
The parameters $s_{\rm th}$ and $f_0$ are the threshold of the continuum and the coupling strength 
between the current and the lowest-lying pole, respectively. 
To make the integral on the right-hand side of Eq.~(\ref{eq:disp0}) 
dominated by the pole contribution and improve the convergence of the OPE, 
we use the Borel transform defined by 
\begin{eqnarray}
\B[f(Q^2)] \equiv \!\!
\lim_{ \substack{Q^2, n \rightarrow \infty \\ Q^2/n = M^2 } }  
\frac{(Q^2)^{n+1}}{n!} \left( - \frac{d \ }{dQ^2 } \right)^n f(Q^2) 
\label{eq:Borel}
.
\end{eqnarray}
Since the right-hand side of Eq.~(\ref{eq:disp0}) is transformed to 
\begin{eqnarray}
\B[{\rm rhs \ of \ Eq.}\,(\ref{eq:disp0})] = \int \!\! \rho(s) e^{-s/M^2} ds, 
\label{eq:afterBorel}
\end{eqnarray}
the contribution from the continuum in the high-energy 
perturbative region is exponentially suppressed. 
By applying the Borel transform to both sides of Eq.~(\ref{eq:disp0}), 
we find the mass formula  
\begin{eqnarray}
m^2 =  - \frac{\partial \ }{ \partial(1/M^2)} \ln[ \M_{\rm OPE} - \M_{\rm cont} ]
\label{eq:mass}
\end{eqnarray}
with 
\begin{eqnarray}
\M_{\rm OPE} &=& \B[ \Pi_{\rm OPE}^J], \\  
\M_{\rm cont} &=& \frac{1}{\pi} \int_{s_{\rm th}}^{\infty} {\rm Im} \Pi_{\rm pert}(s) e^{-s/M^2} ds. 
\label{eq:cont.definition}
\end{eqnarray}
Since the results should be independent of the unphysical Borel parameter $M^2$, 
we should find a window in which 
the mass formula (\ref{eq:mass}) is approximately independent of $M^2$. 
One can estimate the precision of the results 
by the $M^2$ dependence of the mass within the Borel window. 
This is the conventional procedure in QCD sum rule analyses.

We, however, need to reexamine the {\it Ansatz} (\ref{eq:pole}) 
when investigating effects of an externally applied magnetic field. 
Its effects should be consistently taken into account 
on both sides of Eq.~(\ref{eq:disp0}) 
because effects of magnetic fields appear at both hadronic and quark levels. 
In the next section, we show how the {\it Ansatz} (\ref{eq:pole}) should be modified 
in the presence of an external magnetic field, and obtain appropriate {\it Ans\"atze} for 
neutral and charged $D$ mesons, respectively.


\begin{figure*}[t!]
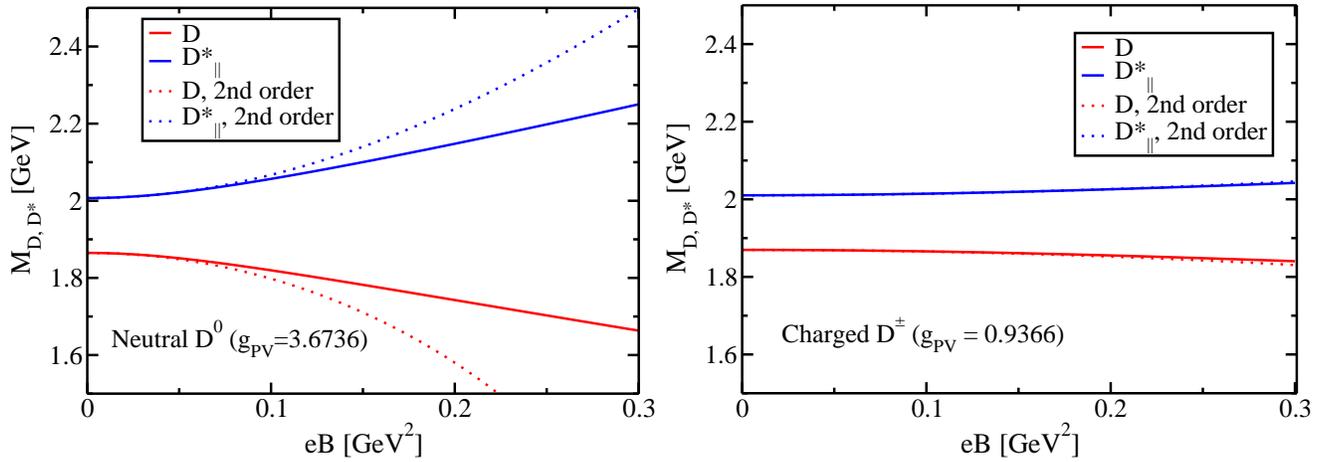

\begin{minipage}[t]{0.48\hsize}
 \begin{center}  
   \includegraphics[clip,width=1\columnwidth]{NeutralD-Dstar_mixing}
    \end{center} 
\end{minipage}
\begin{minipage}[t]{0.48\hsize}
 \begin{center}  
   \includegraphics[clip,width=1\columnwidth]{ChargedD-Dmixing_ver3}
    \end{center} 
\end{minipage}
 \caption{Level repulsion due to a mixing between the $D$ and the longitudinal $D^\ast$ mesons.
The left panel shows masses of neutral $D$ and $D^{*}$, 
while the right panel show those of charged ones.}
 \label{fig:D_mixing}
\end{figure*}

\section{Effects of Hadronic mixing and Landau levels on the phenomenological side}

\label{sec:ph}

In this section, we discuss effects of external magnetic fields on the meson spectrum 
by using effective field theories composed of mesonic degrees of freedom. 
This will be helpful not only to understand the naive expectation of the effects but also to 
construct the appropriate spectral {\it Ansatz} (\ref{eq:pole}) on the phenomenological side of QCDSR. 
In the following, we discuss two important effects, which are 
magnetically induced mixing effects and Landau levels.

\subsection{Mixing effects}
\label{sec:mix}


Effects of magnetic fields give rise to mixing among spin eigenstates \cite{Mac, AS, Bonati:2015dka, Letter, PRD}, such as between 
$\eta_c$ and one spin state of $J/\psi$. 
This mixing is caused by the breaking of a part of the spatial rotation symmetry: 
there remains only the azimuthal component 
along the direction of the external magnetic field. 
This indicates that only the spin state along the magnetic field 
can persist as a good quantum number of the meson systems, and thus that 
there is a mixing between the eigenstates of the total spin, 
$(S_{\rm total}, S_z) = (1,0) $ and $(0,0)$. 
The vector meson state with this polarization 
is called the longitudinal state below. 
Mixing effects will also arise in the spectra of both charged and neutral $D$ mesons. 

In two recent papers \cite{Letter, PRD}, the QCDSR has been extended so that 
hadronic mixing effects are tractable on the phenomenological side. 
An important observation in these works was that the mixing effect induces two adjacent poles 
in the spectral function $\rho(q^2)$ corresponding to the original lowest-lying pole 
and its mixing partner. 
In the present pseudoscalar channel case, these are 
the $D$-meson and the longitudinal $D^\ast$-meson poles, 
and {\it vice versa} in the vector channel. 
To obtain the precise mass spectrum from QCDSR, 
contributions from these poles have to be treated separately, as otherwise, 
one would obtain an average of the two pole masses from a naive QCDSR analysis with a single pole. 
However, it is difficult to separate them by using the Borel transform 
unlike the separation between the lowest-lying pole and the continuum, 
because these poles are close to each other. Moreover, 
the mixing effect becomes stronger as the poles approach each other, 
and the contamination thus becomes a serious problem. 
Hence, it is crucial to have a sufficiently accurate knowledge of the mixing strength 
between the two states.

This mixing strength can in our case 
be estimated by using the effective interaction vertex (see Ref.~\cite{PRD}) 
\begin{eqnarray}
\Lag_{\gam \pv} &=&
\frac{ g_{\pv} }{ m_0 } e \tilde{F}_{\mu \nu} (\partial^{\mu} P) V^{\nu}
\label{eq:L_pv} 
\end{eqnarray}
where $P$ and $V^\mu$ represent the pseudoscalar and vector fields, 
and $F^{\mu\nu}$ and $m_0$ are the field strength tensor of the external magnetic field 
and the average mass $m_0 = (m_\ps + m_\V)/2$. 
Diagonalizing these states, we find the physical $D$ and $D^\ast$ states 
To obtain the coupling constant $g_\pv$ for the charged case, we fit 
the experimentally measured radiative decay width $\Gamma[V \to P + \gamma]$, as shown 
in the first section of Appendix~\ref{sec:g}. 
For the neutral $D$ meson, the same radiative decay width has so far not been directly 
measured. The estimate for the mixing strength can thus not be as precise as for the charged case. 
In Appendix~\ref{sec:g} we therefore use two independent methods to evaluate $g_\pv$ and its 
systematic uncertainty. 

Let us discuss the mass spectra obtained 
by diagonalizing the mass matrix in the presence of the effective vertex (\ref{eq:L_pv}). 
The physical mass eigenvalues are found to be 
\begin{eqnarray}
\hspace{-0.5cm}
m_{D^\ast, D}^{2} 
&=& \frac{1}{2}
\left(
M_{+}^{2} + \frac{ \gamma^{2} }{ m_{0}^{2} } \pm \sqrt{ M_{-}^{4} + 2 \frac{ \gamma^{2} M_{+}^{2} }{ m_{0}^{2} } + \frac{ \gamma^{4} }{ m_{0}^{4} } },
\right)
\label{eq:Mixing_full}
\end{eqnarray}
where $M_{+}^{2} = m_{\rm{p}}^{2} + m_{\rm{v}}^{2}$, $M_{-}^{2} = m_{\rm{v}}^{2} - m_{\rm{p}}^{2}$, 
and $\gamma = g_\pv eB$.
Expanding the right-hand side of Eq.~(\ref{eq:Mixing_full}) up to the second order in $\gamma$, 
we find  
\begin{eqnarray}
m_{D^\ast, D}^{2} &=&
m_{\V,\ps}^{2} \pm \frac{\gamma^2}{M_-^2}
.
\label{eq:Jpsi_2nd}
\end{eqnarray}
Figure~\ref{fig:D_mixing} shows the mass spectra of the physical $D$- and $D^{\ast}$- meson states. 
We see a level repulsion between $D$ and 
the longitudinal $D^\ast$ mesons for all considered cases. 
In particular, in the neutral case, the second-order perturbation breaks down above $eB \sim 0.1$ GeV$^{2}$ 
for $g_\pv = 3.6736$. 

\begin{figure}[t]
 \centering
  \includegraphics[width=0.95\columnwidth]{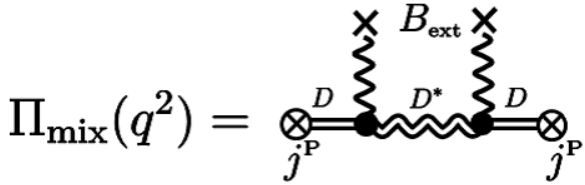}
  \vspace{-0.3cm}
 \caption{Diagrammatic representation of mixing effects from a three-point vertex (\ref{eq:L_pv}).}
 \label{fig:mixing}
\end{figure}

The mixing effect is shown diagrammatically in Fig.~\ref{fig:mixing}, in which 
the $D$ meson created by the pseudoscalar current is mixed with a $D^\ast$ meson 
in the intermediate state through the vertex (\ref{eq:L_pv}). 
Note that we here ignore the direct coupling of the current to the $D^\ast$ meson in the 
presence of a magnetic field. At least in the heavy-quark limit, this coupling was, however, shown to be small~\cite{Letter,PRD}. 
From the diagram in Fig.~\ref{fig:mixing}, we find the spectral {\it Ansatz} 
up to the second order in the external magnetic field as 
\begin{eqnarray}
&&
\label{eq:ansatz_0}
\rho^{(0)} (s) =
\rho_{\rm vac}(s) + \frac{1}{\pi}  {\rm Im} \Pi^{(0)}_\mix (s)
\end{eqnarray}
with 
\begin{eqnarray}
\label{eq:ph_mix}
\Pi_\mix (q^2) &=&
- f_0 \frac{\gamma^2}{M_-^4 }
\biggl[ \, \frac{1}{q^2 - m_{\V}^2} \\
&& \hspace{1.6cm}
- \frac{1}{q^2 - m_{\ps}^2}
- \frac{ M_-^2 }{ (q^2-m_{\ps}^2)^2 }  \, \biggr] . \nonumber
\end{eqnarray}
The superscript of $\rho^{(0)} (s)$ denotes a neutral $D$ meson. 
The roles of the three terms in Eq.~(\ref{eq:ph_mix}) have been elaborated in Refs.~\cite{Letter, PRD}, 
and will be briefly discussed in Sec.~\ref{sec:result}. 

\subsection{Summation of Landau levels}

\label{sec:sum_LL}

For the charged $D$ mesons, we should examine effects of the Landau levels 
in addition to the mixing effect. 
In a magnetic field, the energy levels of spin-0 and spin-1 particles are discretized as 
\begin{eqnarray}
E^{(0)} (n, p_z)&=& \sqrt{m_\ps^2 + (2n +1)eB  + p_z^2}, \label{eq:LLspin0}
\\
E^{(1)} (n, p_z, S_z) &=& \sqrt{m_{\rm{v}}^2 + (2n +1)eB  + p_z^2 + g S_z eB}, \nonumber \\
\label{eq:LLspin1}
\end{eqnarray}
specified by the integer $n \, (\geq0)$ for the Landau levels, 
the continuous momentum $p_z$, and the spin quantum number $S_z$ along the external magnetic field. 
In what follows, we will for simplicity consider the case of zero momentum in the $z$ direction ($p_z=0$) 
and focus on the treatment of the spin-0 particles. Generalizations to nonzero $p_z$ and spin 1 are straightforward. 
Furthermore, we here for the moment ignore possible mixing effects, that were discussed in the previous 
subsection. 

First, it is clear from Eqs.~(\ref{eq:LLspin0}) and (\ref{eq:LLspin1}) 
that in the small-$eB$ limit, higher Landau levels ($n>0$) 
will be located infinitesimally close to the ground state ($n=0$). 
As will be seen in the discussion below, some of their poles, moreover, happen to have negative residues and can therefore not be 
treated as part of the continuum. 
Therefore, the simple ``pole + continuum" {\it Ansatz} 
of Eq.~(\ref{eq:pole}) will 
\text{not} be sufficient to describe the phenomenological side of the sum rules and some method has to be devised 
to take all the Landau levels explicitly into account. 
We will therefore here generalize Eqs.~(\ref{eq:pole})--(\ref{eq:cont.definition}) to the case of an infinite number of 
Landau levels. 

We start by observing that the propagator of a charged spin-0 particle at rest in our conventions changes from 
\begin{eqnarray}
\Pi^{\mathrm{pole}}(s) = - f_0 \frac{1}{s - m_\ps^2 + i \epsilon} 
\end{eqnarray}
in vacuum to 
\begin{eqnarray}
\Pi_{eB}^{\mathrm{pole}}(s) = -2 f_0 \sum_{n=0}^{\infty} (-1)^n \frac{1}{s - m_\ps^2 - (2n+1)|eB| + i\epsilon} \nonumber \\ 
\end{eqnarray}
at finite $|eB|$ \cite{ASPS}. Here, spatial momentum components perpendicular to the magnetic field are taken to be zero. 
Note especially the factor $(-1)^n$, which leads to negative residues for odd $n$ Landau levels. 
Therefore, we have 
\begin{eqnarray}
\frac{1}{\pi} \mathrm{Im} \Pi^{\mathrm{pole}}(s)  = f_0 \delta(s - m_\ps^2) 
\end{eqnarray}
for $|eB|=0$, which is the first term of Eq.~(\ref{eq:pole}), and 
\begin{eqnarray}
\frac{1}{\pi} \mathrm{Im} \Pi_{eB}^{\mathrm{pole}}(s) = 2 f_0 \sum_{n=0}^{\infty} (-1)^n \delta \big(s - m_\ps^2 - (2n+1)|eB| \big)  \nonumber \\
\end{eqnarray}
for nonzero values of $|eB|$, which should replace the pole term in Eq.~(\ref{eq:pole}). 
Let us now compute its contribution to the right-hand side of Eq.~(\ref{eq:afterBorel}). In the vacuum, this gives 
$f_0 e^{-m_\ps^2/M^2}$, which the Landau levels modify as 
\begin{eqnarray}
\int_{0}^{\infty} ds  \frac{1}{\pi} \mathrm{Im} \Pi_{eB}^{\mathrm{pole}}(s) e^{ - \frac{s}{M^2}} 
&=& 2 f_0 \displaystyle \sum_{n=0}^{\infty} (-1)^n e^{- \frac{m_\ps^2 + (2n+1)|eB|}{M^2}} \nonumber \\
&=& 2 f_0  e^{- \frac{m_\ps^2 + |eB|}{M^2}} \frac{1}{1 + e^{-2\frac{|eB|}{M^2}}} \nonumber \\
&=& f_0 e^{-m_\ps^2/M^2} \frac{1}{\cosh \Big( \frac{|eB|}{M^2} \Big)}. 
\label{eq:Landau.integral}
\end{eqnarray}
Comparing this to the vacuum case, we see that, somewhat surprisingly, the effects of the fully resummed Landau levels can be expressed as a simple 
factor $1/\cosh(|eB|/M^2)$. The vacuum formula of Eq.~(\ref{eq:mass}) can thus be easily generalized: 
\begin{eqnarray}
m_\ps^2 &=& - \frac{\partial \ }{ \partial(1/M^2)} \ln \Big[\cosh \big( \frac{|eB|}{M^2} \big) (\M_{\rm OPE} - \M_{\rm cont}) \Big] \nonumber \\
&=& - \frac{\partial \ }{ \partial(1/M^2)} \ln[ \M_{\rm OPE} - \M_{\rm cont} ]  - |eB| \tanh \Big( \frac{|eB|}{M^2} \Big). \nonumber \\
\label{eq:masswithoutLL}
\end{eqnarray}
The quantity $m_\ps^2$ in this formula corresponds to the mass of the charged particle, from which all Landau level 
effects have been subtracted. We emphasize that the above formula is exact and its usage is not restricted to 
small $|eB|$ values. 
In the present work, we will, however, compute the OPE side only up to terms of second order in $|eB|$ and therefore 
for reasons of consistency should retain only terms of the same order in Eq.~(\ref{eq:masswithoutLL}). We 
thus expand the $\tanh$ term to leading order in $|eB|$, which gives 
\begin{eqnarray}
m_\ps^2 =& - \frac{\partial \ }{ \partial(1/M^2)} \ln[ \M_{\rm OPE} - \M_{\rm cont} ]  - \frac{|eB|^2}{M^2}. 
\label{eq:masswithoutLL.expanded}
\end{eqnarray}

\begin{figure}[t]
 \centering
  \includegraphics[width=0.9\columnwidth]{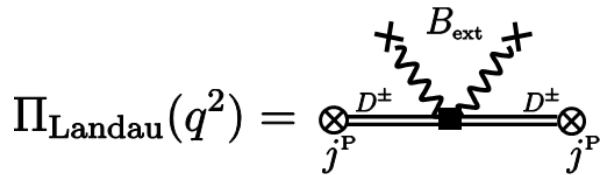}
    \vspace{-0.3cm}
 \caption{A nonvanishing diagram associated with the Landau level of a charged $D$ meson.}
 \label{fig:Landau}
\end{figure}

It is instructive to see 
that the same result can 
be derived perturbatively from the scalar QED effective 
Lagrangian, 
as we shall briefly demonstrate here. We use 
\begin{eqnarray}
\Lag = (D^\mu \phi)^\ast ( D_\mu \phi) - \frac{1}{2} m_\ps^2 \phi^\ast \phi
\, ,
\end{eqnarray}
where the covariant derivative is defined by $D^\mu = \partial^\mu - ie A_\ext^\mu$ 
with an external field $A_\ext^\mu$. 
The above Lagrangian contains linear and quadratic couplings to an external field which are 
proportional to $\phi^\ast A_\ext^\mu \partial_\mu \phi $ and $ \vert \phi \vert^2  A_\ext^2$, respectively. 
Inserting the linear-coupling vertex between two free propagators $D^{(0)} (p) = i /(p^2-m_\ps^2)$, 
we find that it vanishes identically, i.e., $D^{(0)} A_\ext^\mu \partial_\mu D^{(0)} = 0$, 
when using the Fock-Schwinger gauge for an external magnetic field 
$A_\ext^\mu (x) = - 2^{-1} F^{\mu\nu} x_\nu$. 
Therefore, only the quadratic coupling shown in Fig.~\ref{fig:Landau} provides a relevant correction. 
Following a straightforward computation, 
we obtain the second-order correction as
\begin{eqnarray}
\Pi^{\mathrm{pole}}_{eB} (s) 
&=& 
- f_0 \Big[ \frac{1}{s - m_\ps^2 + i \epsilon}
- |eB|^2 \frac{1}{(s - m_\ps^2 +  i \epsilon)^3} \Big], \nonumber \\
\label{eq:Landau}
\end{eqnarray}
where we have as before assumed 
the transverse momentum to be zero. 
Taking the imaginary part and computing the same integral 
as in Eq.~(\ref{eq:Landau.integral}), we get 
\begin{eqnarray}
\int ds \frac{1}{\pi} \mathrm{Im} \Pi_{eB}^{\mathrm{pole}}(s) e^{ - \frac{s}{M^2}} &=& 
f_0 e^{-m_\ps^2/M^2} \Big[1 - \frac{1}{2} \frac{|eB|^2}{M^4} \Big], \nonumber\\
\label{eq:Landau2}
\end{eqnarray}
which coincides with the expanded version of Eq.~(\ref{eq:Landau.integral}) 
up to terms of second order in $|eB|$. 


\section{Operator product expansion for open heavy flavors}

\label{sec:OPE}

\begin{figure*}[tb]
 \centering
 \includegraphics[width=0.8\textwidth]{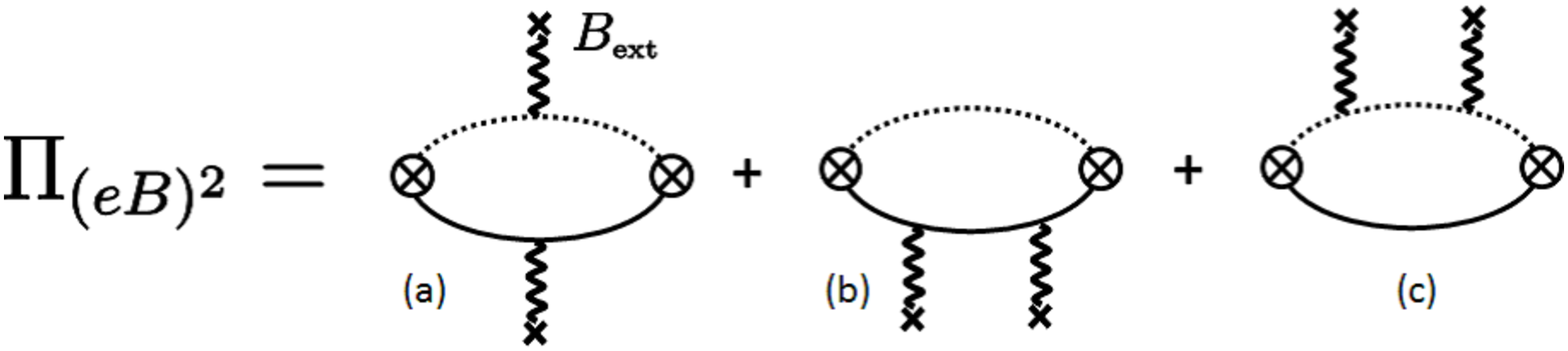}
   \vspace{-0.3cm}
 \caption{
Two insertions of external magnetic fields. 
Dotted and solid lines show light and heavy quarks, respectively.}
 \label{fig:eB2}  
   \vspace{0.5cm}
\centering
 \includegraphics[width=0.9\textwidth]{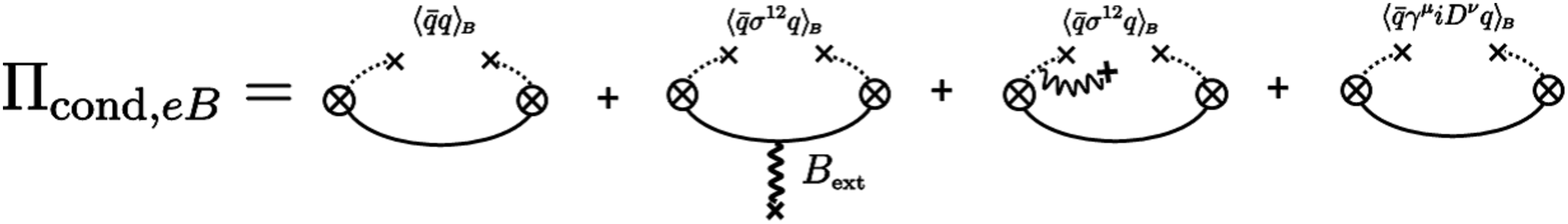}
   \vspace{-0.3cm}
 \caption{Current correlator with quark condensates.}
 \label{fig:cond}
\end{figure*}


In this section, we will discuss the OPE for open heavy flavors in an external magnetic field 
up to dimension-4 operators, which can be divided into three classes 
\begin{eqnarray}
\Pi_{\rm OPE} = \Pi_\vac + \Pi_{(eB)^2} + \Pi_{{\rm cond},\,eB}
\, .
\end{eqnarray} 
The terms commonly included in $\Pi_\vac $ are 
the perturbative terms at leading and next-to-leading orders in $\alpha_s$, 
the scalar gluon condensate $\langle G^{\mu\nu}G_{\mu\nu} \rangle$,  
the light-quark condensate $\langle \bar q q\rangle$, and the mixed light-quark 
condensate $\langle \bar q g \sigma G q\rangle$. 
We use the Wilson coefficients shown in Ref.~\cite{ZHK} for these terms. 
The other terms induced by the external magnetic field, $\Pi_{(eB)^2} $ and $ \Pi_{{\rm cond},\,eB}$, 
are diagrammatically depicted in Figs.~\ref{fig:eB2} and \ref{fig:cond}, respectively, 
and are computed in the subsequent sections.  
We also include effects of the modification of the usual quark condensate $\langle \bar q q \rangle$ due to the 
external magnetic field  
\cite{GMS, IS, ChPT2, qbarq_lat, BMW_qbarq, sus_lat1, sus_lat2, SD2}. 
While the gluon condensate is also modified in a magnetic field, 
we do not take this into account as the modification is estimated to be 
less than 10 \% in the range $eB \lesssim 0.3 \,{\rm GeV}^2 $ \cite{BMW_GG, Ozaki}.


\subsection{External magnetic field insertions}

First, we compute the Wilson coefficients shown in Fig.~\ref{fig:eB2}. 
As explained in Appendix~\ref{sec:eB2app}, we perform the loop integrals 
using the standard 
Feynman parameters, and then take the chiral limit for the light-quark mass. 
The computation in the chiral limit is not only helpful for analytically performing the Borel transform 
but is also necessary for treating infrared singularities and correctly defining the condensates. 
Since there is no momentum transfer from a constant external magnetic field to the quarks, 
the insertions of soft external lines 
induce an infrared divergence in the soft momentum region of 
the loop integral when the chiral limit is taken. This infrared singularity is similar 
to that of the Wilson coefficients for the gluon condensates, 
which is known to be infrared safe due to the cancellation 
with infrared singularities emerging from quark condensates \cite{ZHK,GR84}. 
We show the explicit form of the infrared singularities in this section, 
and find that the OPE in an external constant magnetic field is indeed 
infrared safe in the next section.

For 
comparing our results with the 
corresponding Wilson coefficients of the gluon condensates at finite density \cite{ZHK}, 
we decompose the product of the field strength tensor into 
a scalar part, and a traceless and symmetric tensor part as
\footnote{It is useful to introduce metric tensors 
in the longitudinal and transverse subspaces
$ g_\parallel^{\mu\nu} = {\rm diag} (1,0,0,-1) $ 
and $ g_\perp^{\mu\nu} = {\rm diag} (0, -1, -1, 0)$, respectively. 
With these metrics, we define the longitudinal momentum 
$q_\parallel^\mu = g_\parallel^{\mu\nu} q_\nu = (q^0,0,0,q^3)$ 
and the transverse momentum $q_\perp^\mu = g_\perp^{\mu\nu} q_\nu = (0,q^1,q^2,0)$, 
so that the inner products are given by $p_\parallel \cdot q_\parallel = p^0 q^0 - p^3 q^3 $, 
$p_\perp \cdot q_\perp = - p^1 q^1 - p^2 q^2 $, and so on.}
\begin{eqnarray}
&&
\frac{\alpha_\EM}{\pi} F^{\mu\nu} F_{\mu\nu} = F_0,
\label{eq:F0}
\\
&&
\frac{\alpha_\EM}{\pi} \left( \, F^{\mu\alpha} F^\nu_{\ \, \alpha} 
- \frac{1}{4} g^{\mu\nu} F^{\alpha\beta} F_{\alpha\beta} \right) 
= F_2 (g_\parallel^{\mu\nu} - g_\perp^{\mu\nu}  ), \nonumber \\
&&
\label{eq:F2}
\end{eqnarray}
with $F_0 = \frac{\alpha_\EM}{\pi} 2 B^2 $ and 
$F_2 = \frac{\alpha_\EM}{\pi} \left(-\frac{1}{2} B^2 \right) $. 
Correspondingly, one can decompose the correlator as 
\begin{eqnarray}
\Pi_{(eB)^2} (q) &&= \sum_{ w =a,b,c} 
\left[ \,  \Pi_{(eB)^2}^{(w 0)} (q) + \Pi_{(eB)^2}^{(w 2)} (q) \, \right], \\
\nonumber
\end{eqnarray}
where the superscripts, $a$, $b$, and $c$, represent the corresponding graphs 
with the same index shown in Fig. \ref{fig:eB2}. 
The first and second terms correspond to 
the scalar and the tensor parts proportional to $F_0$ and $F_2$, respectively. 
Following the computation briefly discussed in Appendix~\ref{sec:eB2app}, 
we obtain the scalar part as  
\begin{eqnarray}
\Pi_\eB^{(a0)} (q) &=& - \frac{\Ql \Qh }{8} \kappa F_0 \frac{1}{q^2-\mh^2},
\label{eq:eB_a0}
\\
\Pi_\eB^{(b0)} (q) &=& \frac{Q_h^2 }{24} \kappa F_0 \frac{1}{q^2-\mh^2},
\label{eq:eB_b0}
\\
\Pi _\eB^{(c0)}(q) &=& \frac{\Ql^2 }{24} \kappa F_0 \left( \, 
- 2\frac{\mh}{\ml} \frac{1}{q^2-\mh^2}  + \frac{q^2 - 2 \mh^2}{(q^2-\mh^2)^2} \, \right),
\nonumber
\\
\label{eq:eB_c0}
\end{eqnarray}
and the tensor part as 
\begin{widetext}
\begin{eqnarray}
&& \Pi_\eB^{(a2)} (q) = -Q_l Q_h \kappa F_2 \cdot 3(q_\parallel^2 - q_\perp ^2) \left[ \frac{1}{3q^4} - \frac{1}{6q^2} \frac{1}{q^2-m_h^2} -\frac{1}{3}\frac{m_h^2}{q^6} \ln \left( -\frac{m_h^2}{q^2-m_h^2} \right) \right],
\label{eq:eB_a2}
\\
&& \Pi_\eB^{(b2)} (q) = -Q_h^2 \kappa F_2 \cdot 3(q_\parallel^2 - q_\perp ^2)
\left[ - \frac{1}{9q^4}  - \frac{1}{18q^2} \frac{1}{q^2-m_h^2} + \left( -\frac{1}{9q^4} + \frac{m_h^2}{9q^6} \right) \ln \left( -\frac{m_h^2}{q^2-m_h^2} \right) \right],
\label{eq:eB_b2}
\\
&& \Pi _\eB^{(c2)}(q) = -Q_l^2 \kappa F_2 \cdot 3(q_\parallel^2 - q_\perp ^2)
\left[ - \frac{2}{9q^4}  + \frac{1}{18q^2} \frac{1}{q^2-m_h^2} - \frac{1}{6q^2} \frac{m_h^2}{(q^2-m_h^2)^2} + \left( \frac{1}{9q^4} + \frac{2 m_h^2}{9q^6} \right) \ln \left( -\frac{m_h^2}{q^2-m_h^2} \right) \right. 
\nonumber
\\
&&  \hspace{2.2cm}
\left. - \frac{1}{9q^2} \left( \frac{m_h^2}{(q^2-m_h^2)^2} + \frac{1}{q^2-m_h^2} \right) \ln \left( \frac{m_l^2}{m_h^2} \right) - \frac{2}{9q^2} \left( \frac{m_h^2}{(q^2-m_h^2)^2} + \frac{1}{q^2-m_h^2} \right) \ln \left( -\frac{m_h^2}{q^2 - m_h^2} \right)\right].
\label{eq:eB_c2}
\end{eqnarray}
\end{widetext}
In the above, electric charges and masses of light and heavy quarks are denoted 
as $\Ql$, $\Qh$, $\ml$ and $\mh$, respectively. 
Note that $\Ql \Qh = + 4/9$ for neutral $D$ mesons 
and $\Ql \Qh = - 2/9$ for charged $D$ mesons. 
The constant $\kappa$ is given by traces of the color matrices 
$\kappa = \tr[\id_c]/\tr[t^at^a] = 2N_c$ without summation over $a$, 
which is introduced to compare 
the above expressions 
with the results for the gluon condensates (see Appendix~\ref{sec:eB2app}). 

A few comments are in order. 
(i) The Wilson coefficients (\ref{eq:eB_a0})--(\ref{eq:eB_c2}) contain 
logarithmic and linear infrared divergences in the chiral limit, $\ml \to 0$. 
They are the terms proportional to $1/m_l$ in Eq.~(\ref{eq:eB_c0}) 
and to $\log (m_l^2)$ in Eq.~(\ref{eq:eB_c2}), respectively. 
We will show in the next section that they are exactly canceled by 
divergent terms emerging from quark condensates. 
(ii) We find a simple correspondence between the above expressions 
and those in Eqs.~(C.16)--(C.21) in Ref.~\cite{ZHK} 
computed for the gluon condensates at finite density. 
This correspondence discussed in Appendix~\ref{sec:eB2app} 
indicates consistency between their and our computations. 
However, the results in Ref.~\cite{Machado} do not agree with ours and those given in Ref.~\cite{ZHK}. 
(iii) Based on the unique tensor decomposition shown in Eqs.~(\ref{eq:F0}) and (\ref{eq:F2}), 
we find that the Wilson coefficients obtained above are valid for a general configuration of 
a constant external field $F^{\mu\nu}$, although we have performed the computation 
for a magnetic field oriented in the third spatial direction 
assumed in Eqs.~(\ref{eq:S1}) and (\ref{eq:S2}).


\subsection{Quark condensates}

Next, we compute the Wilson coefficients corresponding to the graphs shown in Fig.~\ref{fig:cond} 
which include light-quark condensates. 
Since quark condensates themselves are perturbatively IR divergent, we need to compute 
not only the Wilson coefficients but also perturbative pieces of the quark condensates to 
extract these divergences at one-loop accuracy as shown in Appendix~\ref{sec:condapp}.

The key technique involved in the perturbative computation of 
the time-ordered current correlator is the contraction of fields 
which are subsequently replaced by Feynman propagators. 
While the residual normal-ordered fields vanish when 
sandwiched between perturbative vacuum states, 
they take finite expectation values in the nonperturbative QCD vacuum. 
Therefore, the time-ordered two-point function contains not only 
a perturbative propagator $S(x) $ 
but also normal-ordered expectation values of bilinear fields as 
$\langle T[  q^a_\alpha (x) \bar q^b_\beta (0) ] \rangle 
= \delta^{ab} S_{\alpha\beta}(x) + \langle :  q^a_\alpha (x) \bar q^b_\beta (0) : \rangle$, 
where superscripts $a$ and $b$ are the color indices 
and subscripts $\alpha$ and $\beta$ the spinor indices. 
Substituting 
the nonperturbative piece ${\mathcal S}^{ab}_{\alpha\beta}(x) 
= \langle :  q^a_\alpha (x) \bar q^b_\beta (0) : \rangle$ 
into the definition of the current correlator (\ref{eq:JJ}), 
we compute the contributions of the condensate terms as 
\begin{eqnarray}
\Pi (q^2) = i \int \!\!\! \frac{d^4p}{(2\pi)^4} 
\tr\left[ {\mathcal S} (p) \gamma^5 S_h (p-q) \gamma^5 \right]
,
\end{eqnarray}
where $S_h(p)$ denotes the heavy-quark propagator with possible insertions of 
external magnetic fields. With the first three 
terms of the propagator, given in  
Eqs.~(\ref{eq:S0})--(\ref{eq:S2}), 
which are ordered in powers of $eB$, we have 
$S_h(p) = S_h^{(0)} (p) + S_h^{(1)} (p) + S_h^{(2)} (p) + \cdots$.

To perform the computation, we first classify the condensates. 
By employing the Fock-Schwinger gauge, specified by $x^\mu A_\mu(x) = 0$, 
the nonperturbative piece in the light-quark propagator 
${\mathcal S}^{ab}_{\alpha\beta}(x) $ can be expanded as 
\begin{widetext}
\begin{eqnarray}
{\mathcal S}^{ab}_{\alpha\beta}(x)
 &=&
\langle : [ \,  q^a_\alpha  (0) + x^\mu D_\mu  q^a_\alpha  (0)
+ \frac{1}{2} x^\mu x^\nu D_\mu D_\nu \ q^a_\alpha  (0) + \cdots \, ] 
\bar q^b_\beta (0) : \rangle 
\nonumber
\\
&=& 
\langle : q^a_\alpha  (0)  \bar  q^b_\beta (0) : \rangle 
+
x^\mu  \langle :  D_\mu q^a_\alpha  (0) \bar  q^b_\beta (0)  : \rangle 
+
\frac{1}{2} x^\mu x^\nu  \langle : D_\mu D_\nu  q^a_\alpha  (0) \bar  q^b_\beta (0): \rangle 
+ \cdots
\label{eq:Sl}
,
\end{eqnarray}
\end{widetext}
where the covariant derivatives contain both QED and QCD gauge fields. 
In the above expansion, we find three types of condensates in the second line. 
The spinor indices of these condensates can be decomposed 
by the complete set $1$, $\gamma^5$, $\gamma^\mu $, $\gamma^5 \gamma^\mu  $, 
and $\sigma^{\mu\nu} = i/2 \cdot [\gamma^\mu, \gamma^\nu] $ as usual. 
Within this decomposition, to extract contributions proportional to external fields, 
Lorentz indices of nonvanishing components must be represented 
by combinations of the field strength tensors 
$F^{\mu\nu}$, $\tilde F^{\mu\nu} = 2^{-1} \epsilon^{\mu\nu \sigma\rho}F_{\sigma\rho}$, and 
$F^{\mu \sigma} F^{\nu}_{\ \, \sigma} - g^{\mu\nu} F^{\sigma\rho}F_{\sigma\rho} /4$. 
Picking up those satisfying discrete symmetries, not vanishing in the chiral limit, 
and being of second order in $eB$, we find the condensates up to dimension 5 
appearing in Eq.~(\ref{eq:Sl}) as 
\begin{eqnarray}
&&
\langle : q^a_\alpha  (0)  \bar q^b_\beta (0) : \rangle =
- \frac{1}{4N_c} \delta^{ab} \langle : \bar q  (0)  q (0) : \rangle
\\
&& \hspace{3cm}
- \frac{1}{8N_c} \delta^{ab}
\langle : \bar q (0) \sigma_{\mu\nu}  q(0) : \rangle (\sigma^{\mu\nu})_{\alpha\beta} ,
\nonumber
\\
&&
x^\mu \langle : D_\mu q^a_\alpha  (0)   \bar  q^b_\beta (0)  : \rangle 
\nonumber
\\
&&  \hspace{1cm} =
- \frac{1}{4N_c} \delta^{ab} x^\mu  \langle : \bar q  (0) \gamma_\nu D_\mu  q (0) : \rangle 
(\gamma^\nu)_{\alpha\beta} ,
\\
&&
\frac{1}{2} x^\mu x^\nu \langle :   D_\mu D_\nu q^a_\alpha  (0) \bar  q^b_\beta (0): \rangle 
\nonumber
\\
&&  \hspace{1cm} =
- \frac{1}{64N_c} \delta^{ab} x^2 e F_{\mu\nu}
\langle : \bar q (0) \sigma^{\mu\nu} q(0) : \rangle \delta_{\alpha\beta} 
.
\end{eqnarray}
Figure.~\ref{fig:cond} shows nonvanishing combinations of the light-quark condensates 
and external magnetic fields emanating from the heavy-quark propagator.

The expectation values of the normal-ordered fields can be 
related to the quark condensates defined with non-normal-ordered fields. 
This computation is carried out in Appendix~\ref{sec:condapp} at the perturbative level 
so that we can extract the infrared divergences of the quark condensates. 
By using the ``redefined'' condensates shown 
in Eqs.~(\ref{eq:redef1}), (\ref{eq:redef2}), and (\ref{eq:redef3_ST}), 
we obtain the quark condensate contributions to the OPE as 
\begin{widetext}
\begin{eqnarray}
\Pi^{\langle \bar q q \rangle}(q^2) 
= 
&&
\langle \bar q q \rangle \frac{m_h}{q^2 - m_h^2} 
+ \frac{1}{12} Q_l^2 \frac{N_c (eB)^2}{\pi^2} \frac{m_h}{m_l}  \frac{1}{q^2 - m_h^2},  
\label{eq:condcontr1}
\\
\Pi^{\langle \bar q \sigma_{12} q \rangle}(q^2) 
=
&& 
(eB) \langle \bar q \sigma_{12} q\rangle \frac{m_h}{(q^2 - m_h^2)^2} 
\Big[Q_h + 
 Q_l  \frac{m^2_h}{q^2-m_h^2} \Big], 
\label{eq:condcontr2}
\\
\Pi^{\langle \bar q \gamma^{\mu} iD^{\nu} q \rangle}(q^2)  
=
&&
 - \langle \bar q \gamma^{\mu} iD^{\nu} q \rangle  
\Big[g_{\mu \nu} \frac{1}{q^2 - m_h^2} -2 \frac{q_{\mu} q_{\nu}}{(q^2 - m_h^2)^2} \Big] 
\nonumber
\\
&& 
-\frac{1}{12}  Q_l^2 \frac{N_c (eB)^2}{\pi^2}  \Bigg[ \frac{1}{q^2 - m_h^2} +  (q^2 +2 q_{\perp}^2) 
\log \Big( \frac{\mu^2}{m_l^2} \Big) \frac{1}{(q^2 - m_h^2)^2} \Bigg]. \label{eq:condcontr3}
\end{eqnarray}
\end{widetext}
One should observe that the IR-divergent pieces above 
\big[the terms proportional to $1/m_l$ in Eq.~(\ref{eq:condcontr1}) and to $\log (m_l^2)$ 
in Eq.~(\ref{eq:condcontr3})\big] have exactly 
the same forms as those in the Wilson coefficients (\ref{eq:eB_c0}) and (\ref{eq:eB_c2}). 
After the cancellation of these divergent pieces, 
we find an IR-finite OPE of the current correlator. 
Note that to obtain a finite result it is necessary 
to include the quark condensate $\langle \bar q \gamma^{\mu} iD^{\nu} q \rangle $, 
which has not been done in Ref.~\cite{Machado}. 
Note also that the quark condensate $\langle \bar q \sigma^{12} q \rangle$ 
has nonvanishing Wilson coefficients as shown in Eq.~(\ref{eq:condcontr2}). 
These come from the second and third diagrams in Fig.~\ref{fig:cond}, 
which contain both an external field line as well as a quark condensate. 
Whereas the Wilson coefficient vanishes without external field lines as mentioned in Ref.~\cite{Machado}, 
one should include these diagrams as well 
to consistently perform the OPE up to ${\mathcal O}\left( (eB)^2 \right)$, 
since this condensate is proportional to $eB$ in the weak-field limit. 
Finally, one should not explicitly include the heavy-quark condensate $\langle \bar Q_H Q_H \rangle$ 
because this contribution is infrared finite and implicitly included 
as part of the Wilson coefficients for 
the gluon condensate term in the heavy-quark limit \cite{RRYrev}. 

In the next section, we show the mass spectra from QCDSR analyses 
using the OPE obtained above up to dimension-4 operators 
and second order in $eB$.


\section{Results and discussions}

\label{sec:result}

In this section, we first show the mass formulas, 
and then discuss results of the QCD sum rule analyses separately for neutral and charged $D$ mesons.

\subsection{Mass formulas}

We compute the mass spectra of the neutral and charged $D$ mesons by combining 
the spectral {\it Ans\"atze} and the OPE obtained in Secs.~\ref{sec:ph} and \ref{sec:OPE}, respectively. 
With these pieces, we derive the mass formula for the neutral $D$ mesons as 
\begin{eqnarray}
m^2_{ \Dn } =  - \frac{\partial \ }{ \partial(1/M^2)} \ln[ \M_{\rm OPE} - \M_{\rm cont}  -  \M_{\rm mix} ]
\label{eq:mass_0}
\, ,
\end{eqnarray}
where $\M_{\rm OPE} $, $ \M_{\rm cont}  $, and $  \M_{\rm mix}$ 
are the Borel-transformed forms of the OPE, the continuum term, and the mixing term, respectively. 
The last two terms are from the phenomenological side. 
For the $\M_{\rm OPE} $, the Borel-transformed Wilson coefficients 
are summarized in Appendix.~\ref{sec:Borel}. 
The mixing term (\ref{eq:ph_mix}) on the phenomenological side is transformed to be 
\begin{eqnarray}
\M_{\rm mix} = f_0 (eB)^2 \frac{g_{\rm PV}^2}{M_-^4} \left(  e^{- \frac{m_\mathrm{V}^2}{M^2}} - e^{- \frac{m_\mathrm{P}^2}{M^2}} + \frac{M_-^2}{M^2} e^{- \frac{m_\mathrm{P}^2}{M^2}} \right)
,
\nonumber
\label{eq:mix_Borel}
\\ 
\end{eqnarray}
where the terms in the above are shown 
in the same order as in Eq.~(\ref{eq:ph_mix}).

Due to the effects of the Landau levels for the charged $D$ mesons, 
the mass formula is changed according to Eq.~(\ref{eq:Landau2}) as 
\begin{eqnarray}
m^2_{ \Dpm } =&&  - \frac{\partial \ }{ \partial(1/M^2)} \ln[ \M_{\rm OPE}  
- \M_{\rm cont} -  \M_{\rm mix} - \M_{\rm LL}], \nonumber\\
\label{eq:mass_pm}
\end{eqnarray}
where $\M_{\rm LL}$ is the Borel-transformed form of the Landau level term:
\begin{eqnarray}
\M_{\rm LL} = - \frac{1}{2} f_0 e^{-m_\ps^2/M^2} \frac{|eB|^2}{M^4}.
\label{eq:LL_Borel}
\end{eqnarray}
We have ignored the effects of the Landau levels on the mixing terms 
as they are of higher order in $|eB|$.
While this term was also included in Ref.~\cite{Machado}, our interpretation for the role of this term, which we will propose below,
is different from what has been discussed there.
The summation of Landau levels was discussed in Sec.~\ref{sec:sum_LL}. 
To be consistent with the second-order mixing terms of Eq.~(\ref{eq:mix_Borel}), 
we have in Eq.~(\ref{eq:mass_pm}) above retained only terms up to second order in $|eB|$. 

As input parameters, we use the charm-quark pole mass $m_c=1.6 \,{\rm GeV}$, the strong coupling constant $\alpha_s=0.3$ and condensates up to dimension-5 in 
vacuum, $\langle \bar{q} q \rangle_\mathrm{vac}=(-0.27)^3 \mathrm{GeV}^3$~\cite{Aoki:2013ldr}, $\langle \frac{\alpha}{\pi} G^2 \rangle_\mathrm{vac}=0.33 \,{\rm GeV}^4$~\cite{SVZ792}
and $\langle \bar{q} g \sigma G q \rangle_\mathrm{vac} = 0.63 \cdot \langle \bar{q} q \rangle_\mathrm{vac} \,{\rm GeV}^5 $
\footnote{The charm-quark pole mass is $m_c=1.67 \,{\rm GeV}$ at $\mu=m_c$~\cite{PDG}.
The conventional value of the mixed condensate is $\langle \bar{q} g \sigma G q \rangle_\mathrm{vac} = 0.8 \cdot \langle \bar{q} q \rangle_\mathrm{vac} \,{\rm GeV}^5 $ at $\mu=1 \, \mathrm{GeV}$ as estimated in Ref.~\cite{Belyaev:1982sa}.
These values are run to $\mu=2 \,{\rm GeV}$ by using the renormalization group equation.
The anomalous dimension of $\langle \bar{q} g \sigma G q \rangle_\mathrm{vac}$ at one-loop level is given by Ref.~\cite{Narison:1983kn}}.
Here, $m_c$, $\alpha_s$, $\langle \bar{q} q \rangle_\mathrm{vac}$ and $\langle \bar{q} g \sigma G q \rangle_\mathrm{vac}$ are renormalized at 
$\mu=2 \,{\rm GeV}$ to match the renormalization points with those of the lattice QCD observables. 
Furthermore, we employ the following values of the condensates in a magnetic field: 
For the $B$--dependence of $\langle \bar q q \rangle$, we use a combination of chiral perturbation theory \cite{ChPT2} and lattice QCD results \cite{BMW_qbarq}. 
For $eB \leq 0.1 \mathrm{GeV}^2$, the average value of $\langle \bar{u} u \rangle$ and $\langle \bar{d} d \rangle$ obtained in Ref. \cite{ChPT2} agrees with 
that of Ref. \cite{BMW_qbarq} so that we can utilize the analytic representation of Ref. \cite{ChPT2}. 
On the other hand, for $eB > 0.1 \mathrm{GeV}^2$, we make use of the lattice QCD results \cite{BMW_qbarq}, for which the modifications of $\langle \bar{u} u \rangle$ and $\langle \bar{d} d \rangle$ in a magnetic field turn out to be different.
Furthermore, when the magnetic field is weak enough, the $\langle \bar{q} \sigma_{12} q \rangle$ condensate increases linearly with the magnetic field so that we can write
\begin{equation}
\langle \bar{q} \sigma_{12} q  \rangle_B = Q_l (eB) \langle \bar{q}_l q_l  \rangle \chi_l \equiv Q_l (eB) (-\tau_l).
\end{equation}
Thus, the tensor quark condensate $\langle \bar{q} \sigma_{12} q \rangle_B$ 
depends on the light-quark flavor. $\chi_l$ is called magnetic susceptibility 
of the light-quark condensate and $\tau_l$ is the tensor coefficient of the magnetic susceptibility. 
Here, we use $\tau_u = -0.0407\,\mathrm{GeV}$ and $\tau_d = -0.0394\,\mathrm{GeV}$ calculated by lattice QCD \cite{sus_lat2} \footnote{In the work of Ref.~\cite{sus_lat2},  $\sigma_{\mu \nu}=(1/2i) [\gamma_{\mu}, \gamma_{\nu}]$ was used instead of our $\sigma_{\mu \nu}=(i/2) [\gamma_{\mu}, \gamma_{\nu}]$. 
Therefore, we have changed the sign of $\tau_l$ obtained in \cite{sus_lat2} 
to be consistent with our conventions.}. 
Finally, $\langle \bar q \gamma^{\mu} iD^{\nu} q \rangle_B$ is estimated using a simple constituent quark model in Appendix.~\ref{sec:simplemodel}. 
On the phenomenological side, we employ $f_0 \equiv f_D^2 m_{\rm P}^4/m_c^2 = 0.2060 \,{\rm GeV}^4$, $f_D = 0.2046 \,{\rm GeV}$, 
$m_{\rm P} = 1.884\,{\rm GeV}$, $m_{\rm V} = 2.026\,{\rm GeV}$, where $m_{\rm P}$ is the value obtained from our sum rule and $m_{\rm V}$ is 
rescaled by $m_{\rm P}$ and the ratio of the experimental values $m_{D^\ast}/m_D$. 
The value of $g_{\rm PV}$ is estimated in Appendix.~\ref{sec:g}.

In general, hadron masses extracted from sum rules have a Borel mass dependence.
In order to obtain accurate results, 
we determine an effective range of the Borel mass, the so-called Borel window. 
The lower limit of the Borel window can be chosen such that the absolute value of the highest-order $ \langle \bar{q} g \sigma G q \rangle $ term is less 
than $30 \%$ that of $ \langle \bar{q} q \rangle $ term, which happens at about $M_\mathrm{min} \approx 0.87\,{\rm GeV}$. 

\begin{figure*}[t!]
\begin{minipage}[t]{0.48\hsize}
 \begin{center}  
   \includegraphics[clip,width=1\columnwidth]{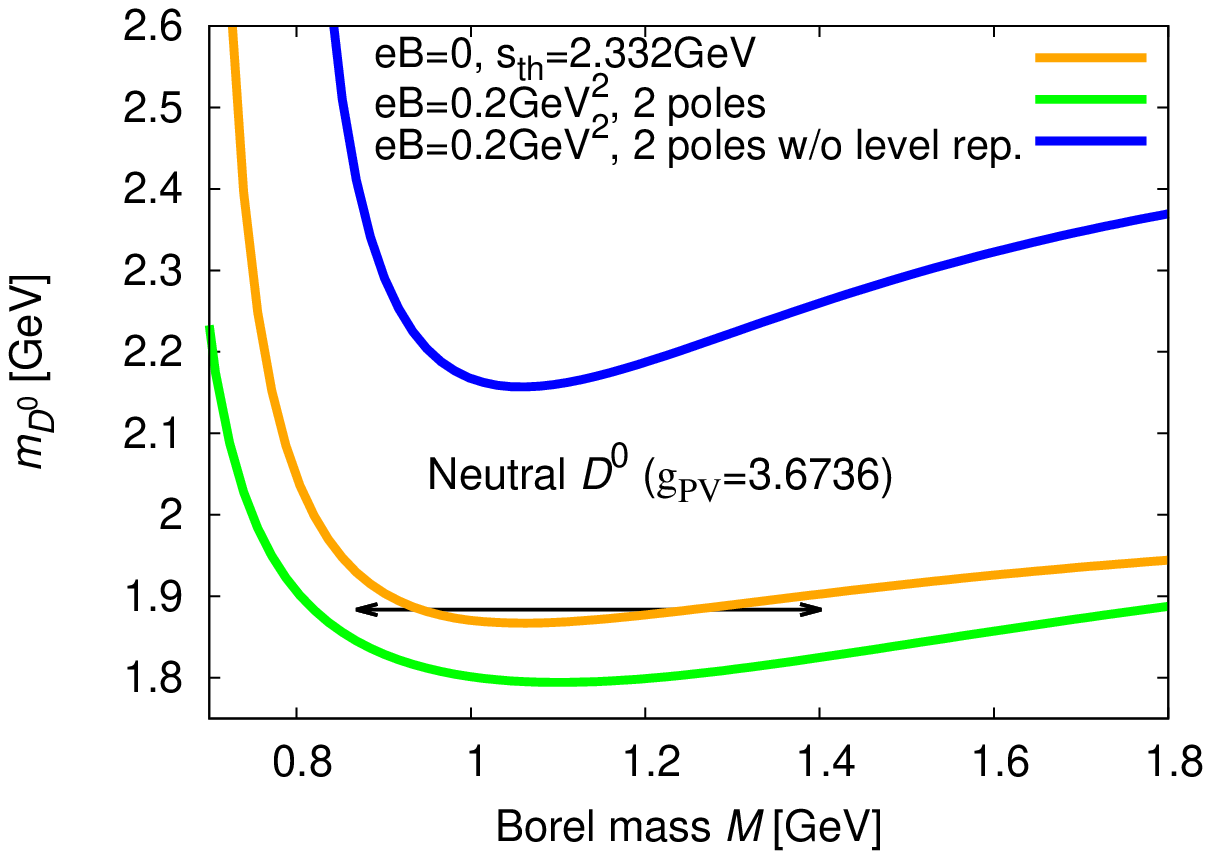}
    \end{center} 
\end{minipage}
\begin{minipage}[t]{0.48\hsize}
 \begin{center}  
   \includegraphics[clip,width=1\columnwidth]{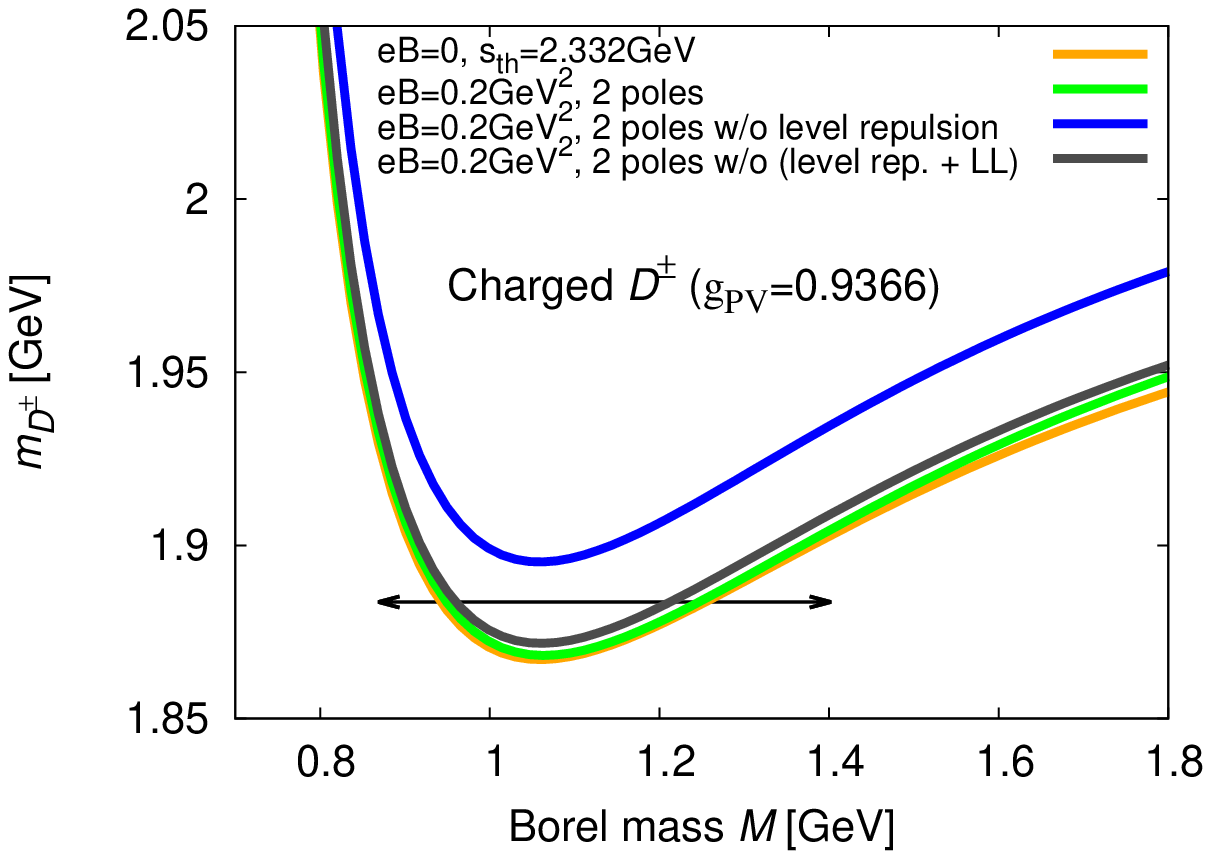}
    \end{center}
\end{minipage}
 \caption{Dependences of the neutral and charged $D$-meson masses, $m_{D^0}$ and $m_{D^\pm}$, on the Borel mass $M$, calculated from 
Eqs.~(\ref{eq:mass_0}) and (\ref{eq:mass_pm}) in the vacuum and for a nonzero magnetic field. 
Black arrows indicate the location of the Borel window determined from OPE convergence and lowest pole dominance.}
 \label{fig:Borel.curve}
\end{figure*}

The OPE above $M_\mathrm{min}$ can then be assumed to be sufficiently convergent because the contribution of the dimension-5 condensate is reasonably smaller than that of the dimension-3 condensate, which gives the largest contribution to the OPE.
On the other hand, the upper limit is determined by requiring the lowest pole contribution to be more than $50\%$ of the total OPE. 
This value is estimated as $M_\mathrm{max} \approx 1.19 \mathrm{GeV}$, and turns out to be almost $B$ independent. 
Once the Borel window is determined, we fix the threshold parameter such that the Borel mass dependence of 
Eqs.~(\ref{eq:mass_0}) and (\ref{eq:mass_pm}) becomes minimal.
A few typical Borel curves are shown in Fig.~\ref{fig:Borel.curve} for the vacuum and nonzero magnetic field. 

\begin{figure*}[t!]
\begin{minipage}[t]{0.48\hsize}
 \begin{center}  
   \includegraphics[clip,width=1\columnwidth]{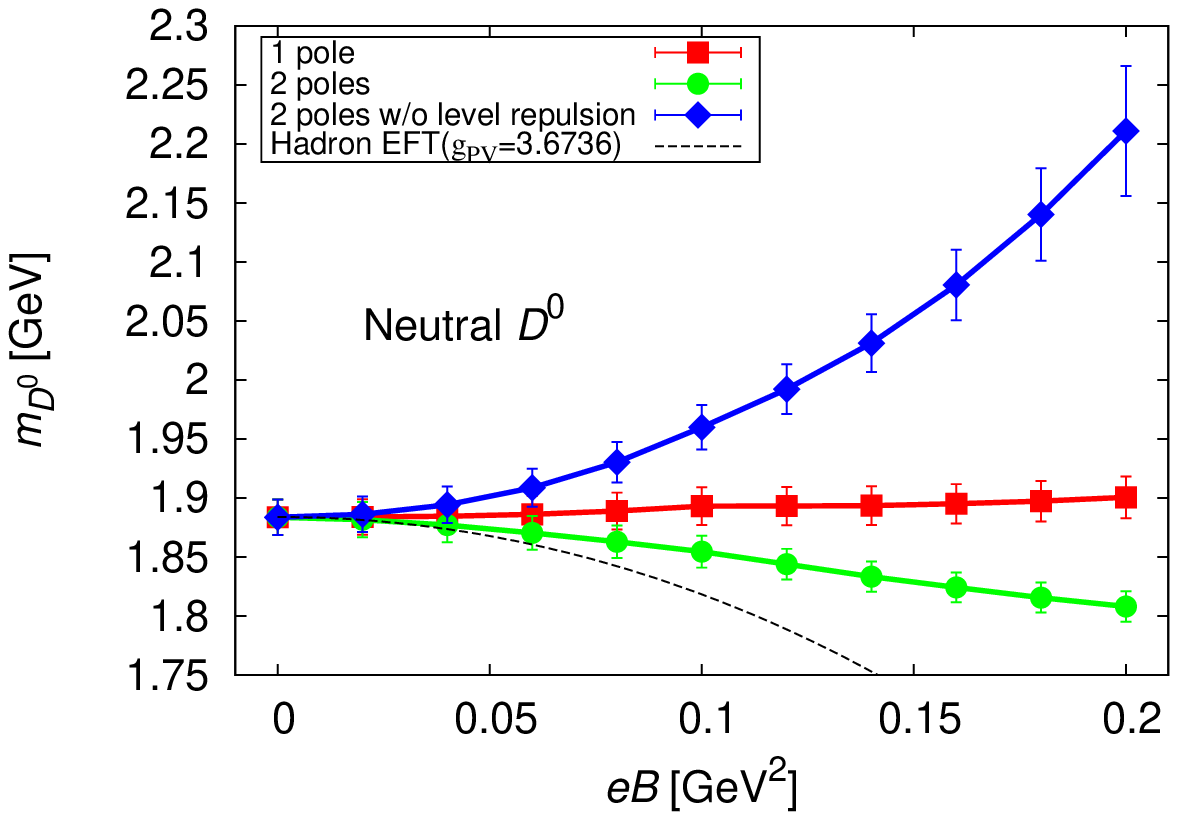}
    \end{center} 
    \vspace{-0.5cm}
     \caption{Mass shifts of neutral $D$ mesons in a weak magnetic field from QCD sum rules. 
The dashed line is estimated from Eq.~(\ref{eq:Jpsi_2nd}) and $g_{\rm PV}=3.6736$.}
\label{fig:result_N}
\end{minipage}
\hspace{0.2cm}
\begin{minipage}[t]{0.48\hsize}
 \begin{center}  
   \includegraphics[clip,width=1\columnwidth]{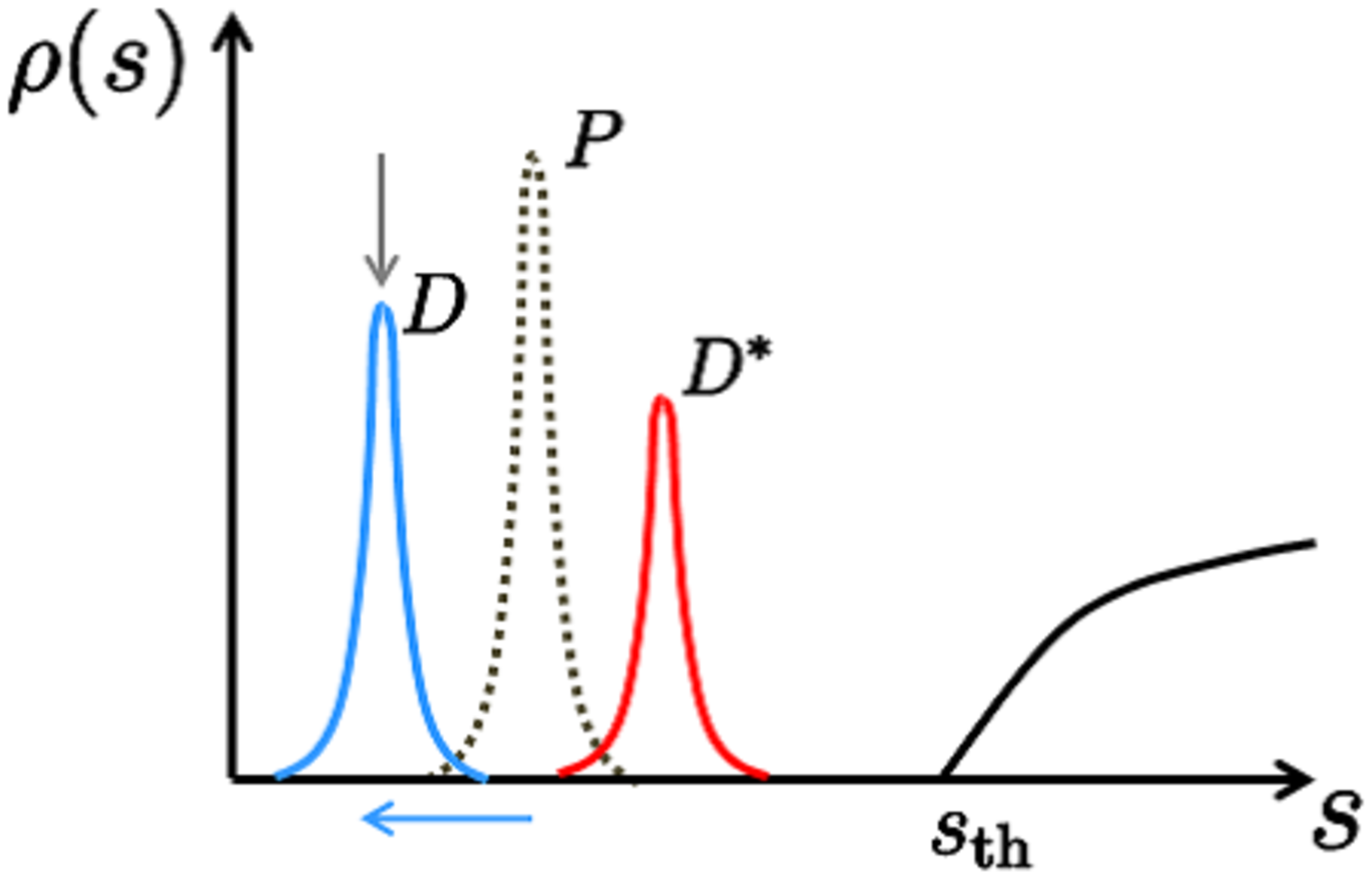}
    \end{center} 
    \vspace{-.5cm}
     \caption{Sketch of the spectral function in the neutral pseudoscalar channel in the presence of the mixing effect. 
The vacuum $D$-meson pole is denoted as a dotted line and marked with $P$ (see also Sec.~\ref{sec:mix}).
Mixing effects lead not only to the appearance of the $D^\ast$-meson pole, 
but also to a reduction of the $D$-meson pole residue and to its shift due to level repulsion. 
These effects are indicated as gray and blue arrows, respectively. 
}
 \label{fig:spectralF}
\end{minipage}
\end{figure*}

It is seen in these figures that the Borel window is wide and that the Borel mass dependence of the physical masses is reasonably small. 
Finally, we take the average values of the mass formulas within this Borel window.
These are plotted as a function of  $eB$ in Figs.~\ref{fig:result_N} and \ref{fig:result_C}, 
together with the systematic errors. 
The error bars correspond to the standard deviations estimated from averaging the Borel curve within the Borel window.

\subsection{Neutral $D$ meson}

We first discuss the results for the neutral $D$-meson spectrum shown in Fig.~\ref{fig:result_N}. 
At the hadronic level, we anticipate only mixing effects without Landau levels, 
because, while quarks appearing on the OPE side are charged, 
the total charge in the neutral $D$-meson system is zero. 
Other effects, that cannot be described by hadronic degrees of 
freedom, if present, will be observed as deviations from the spectrum with the mixing effect. 
For charmonia \cite{Letter, PRD}, effects of a weak magnetic field 
were found to be almost saturated by mixing.

The final result for the neutral $D$ meson is shown by the green line in Fig.~\ref{fig:result_N}, 
which is compared to two different analyses below. 
To get the green line, two poles should be taken into account on the phenomenological side 
since an additional $D^\ast$-meson pole is induced by the mixing effect 
as sketched in Fig.~\ref{fig:spectralF}. As shown in this figure, mixing causes not only 
the appearance of the $D^\ast$-meson pole, but also the reduction of the residue 
of the original $D$ meson and its shift due to level repulsion. 

Let us discuss the roles of the terms 
implemented on the phenomenological side of Eq.~(\ref{eq:mass_0}). 
Including the $D^\ast$-meson pole shown in the first term in Eq.~(\ref{eq:mix_Borel}), 
we subtract the corresponding information from the OPE in Eq.~(\ref{eq:mass_0}). 
One should also include the second term in Eq.~(\ref{eq:mix_Borel}) 
which takes care of the decrease of the residue of the $D$-meson pole. 
Note that this is the reason why this term has a minus sign. 
After the subtraction of the $D^\ast$-meson pole 
and the modification of the $D$-meson residue, the $D$-meson spectrum 
is properly extracted without contamination of the $D^\ast$ pole. 
The green line is obtained by including these two terms, without the last term 
which will be discussed shortly. 
To illustrate the importance of this subtraction and to estimate a magnitude of the contamination, 
we compare the green line with the analysis performed without 
including any term in Eq.~(\ref{eq:mix_Borel}). 
The result of this analysis is shown by a red line. 
In this case, we find a larger mass because this mass is, roughly speaking, 
an average of the $D$- and $D^\ast$-meson masses. 
We find that the deviation from the green line is significant.

Finally, we discuss the last term in Eq.~(\ref{eq:mix_Borel}). 
The physical consequence of the double-pole structure of this term, shown in Eq.~(\ref{eq:ph_mix}), 
is the mass shift caused by the level repulsion, which can be diagrammatically understood as follows.  
In Fig.~\ref{fig:mixing}, the current excites an on-shell $D$-meson state, 
which is then virtually excited to an off-shell $D^\ast$-meson state 
by an interaction with the magnetic field 
and comes back again to an on-shell $D$-meson state. 
Therefore, these two on-shell states correspond to the double pole, 
and the mixing with the off-shell state gives rise to a mass shift. 
Including the last term in Eq.~(\ref{eq:mix_Borel}), 
the level repulsion effect, which is implicitly encoded in the OPE at the quark level, 
is canceled by the corresponding term explicitly implemented at the hadronic level. 
This is clearly seen as a subtraction in Eq.~(\ref{eq:mass_0}). 
Thus, including all terms in Eq.~(\ref{eq:mix_Borel}), 
one would expect to obtain the $D$-meson vacuum mass 
if effects of the magnetic field would be saturated by mixing effects. 
The analysis result taking into account all three terms is shown by the blue line. 
A significant and increasing deviation from the vacuum mass is observed 
for $eB$ values larger than about $0.1\,\mathrm{GeV}^2$. 
We can therefore conclude that some other effect that cannot be 
expressed by hadronic degrees of freedom, which leads to a positive mass shift, 
must exist. This is not surprising, as at $eB = 0.1\,\mathrm{GeV}^2$, 
we are already at scales comparable to $\Lambda_{\mathrm{QCD}}$, which means 
that the magnetic field is strong enough to probe and modify the 
internal structure of hadrons, governed by QCD. 
Our OPE should still work in this regime since the Borel mass, 
serving as the separation scale, is roughly 1~GeV or larger as seen in Fig.~\ref{fig:Borel.curve}.
The detailed mechanisms of the above-mentioned nonhadronic effect will have to be 
clarified in the future.

\subsection{Charged $D$ meson}

\begin{figure}[t]
 \centering
  \includegraphics[clip,width=0.95\columnwidth]{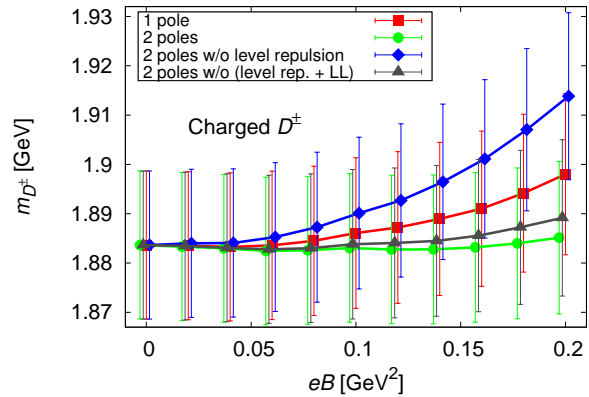}
 \vspace{-0.3cm}
\caption{Mass shifts of charged $D$ mesons in a weak magnetic field from QCD sum rules.
}
\label{fig:result_C}
\end{figure}

Next, we show results for the spectrum of the charged $D$ meson in Fig.~\ref{fig:result_C}. 
As for $D^0$, the charged $D$-meson spectrum is also 
modified by the mixing effect in magnetic fields. 
Therefore, we include the mixing terms in Eq.~(\ref{eq:mix_Borel}) 
whose roles have been already discussed above. 
The mixing strength for the charged $D$ meson is estimated in Appendix~\ref{sec:g}.

In Fig.~\ref{fig:result_C}, a green line shows the mass spectrum obtained 
with the subtraction of the charged $D^\ast$-meson pole as in the previous subsection. 
Again, the difference between the green and blue lines shows 
the magnitude of the mixing effect.
The magnitude is found to be smaller than that for the neutral $D$ meson. 
This can be understood from the fact that the electric charge of the $d$ quark is 2 times 
smaller than that of the $u$ quark, so that we naively expect the mixing effect to be reduced 
by a factor of 4 at second order in $eB$. 
It should be mentioned here that the 
uncertainty of the mixing strength for the charged $D$ meson is smaller than 
that for its neutral counterpart, because the experimental data 
for the radiative decay width of charged $D$ mesons is available.

While the green line shows the mass in the presence of the Landau levels, 
one should note that this mass can {\it not} be interpreted as 
the spectrum of the ground state, i.e., the lowest Landau level. 
The reason is that an infinite number of Landau levels appear in the spectral density 
of the charged pseudoscalar channel as discussed in Sec.~\ref{sec:sum_LL}, 
and the intervals between the adjacent Landau levels are too narrow 
to be resolved in the weak-field limit [cf. the dispersion relation (\ref{eq:LLspin0})]. 
Therefore, while it would be possible to extract the ground state in the strong-field limit 
where the lowest Landau level is isolated from all the higher ones, 
the same strategy does not work in the weak-field limit. 
This fact has been overlooked in Ref. \cite{Ayala}.

As we work here in the weak-field limit, we do not try to extract the lowest Landau level, 
but instead sum up all of them on the phenomenological side 
as given in Eq.~(\ref{eq:LL_Borel}). 
This term subtracts the effects of the Landau levels from the OPE, 
so that the result with this subtraction 
will not include them. 
We will hence interpret our results as follows. 
If the mass spectrum with the above subtraction agrees with the vacuum mass, 
one concludes that effects of the magnetic field are saturated 
by the mixing effect and the Landau levels. 
Otherwise, i.e., if it does not agree with the vacuum mass, 
one would observe residual effects not described by hadronic degrees of freedom. 
On this point, we differ fundamentally from the interpretation put 
forward in Ref. \cite{Machado}, where Landau levels were claimed to be 
observed even though the same term as ours was subtracted on the 
phenomenological side.

The gray line in Fig.~\ref{fig:result_C} shows the result with the subtractions of both Landau levels 
and mixing effects, so that this line is expected to be flat 
if there is no residual effect. 
As can be seen in the figure, we find that the gray line is flat within the error bars, 
and conclude that, in contrast to the neutral case, effects of the magnetic field are consistent with 
the hadronic picture---that is, mixing and Landau levels---in the weak-field limit.

\section{Summary}

\label{sec:summary}

We investigated neutral and charged $D$-meson spectra in the presence of 
external magnetic fields by using QCD sum rules, 
and have established a basic framework for investigating hadrons containing light quarks. 
On the phenomenological side, we carefully examined the spectral {\it Ansatz}  
so that the magnetically induced mixing effects and Landau levels, 
expected at the hadronic level, are properly taken into account. 
Summation of the Landau levels for the spectral function of charged $D$ mesons 
was performed with the help of the Borel transform. 
These treatments are crucial for consistent QCD sum rule analyses in magnetic fields, 
since magnetic fields act on charged mesons on the phenomenological side 
as well as quarks on the OPE side. 
On the OPE side, we implemented magnetic field effects up to the dimension-4 operators 
and terms up to $|eB|^2$. 
We also compared our result with those in Refs.~\cite{ZHK, Machado} and 
explicitly demonstrated the cancellation of infrared singularities. 
Values of the condensates in magnetic fields were estimated 
by means of lattice QCD results where available ($\langle \bar q  q \rangle$, 
$\langle \bar q \sigma^{\mu\nu} q \rangle$), and otherwise 
by a simple model on the basis of constituent quarks ($\langle \bar q  \gamma^{\mu} iD^{\nu} q \rangle$).

By using the framework constructed in this work, 
we discussed the mass spectra of neutral and charged $D$ mesons. 
For neutral $D$ mesons, we found that besides mixing, there must be 
an additional effect of nonhadronic origin, which leads to a 
positive mass shift for sufficiently strong, yet perturbative, magnetic fields. 
On the other hand, the mass spectra of charged $D$ mesons turned out 
to be saturated by mixing and Landau level effects 
within the precision of the present calculation. 
The uncertainties of our approach mainly come from the estimate of the mixing strengths, 
and the not-well-constrained condensates of dimension 5 and higher.

To look for more pronounced effects originating from, 
e.g., modification of the vacuum structure in the presence of magnetic fields, 
we need more precise calculations 
and/or should proceed to the strong-field limit, where the vacuum structure and 
the internal structure inside mesons are more strongly modified. 
Also, the coupling strengths in the heavy-light systems are 
known to be enhanced by the Kondo effect emerging under certain environments. 
Indeed in QCD, recent renormalization group studies have shown 
the strong coupling nature of the mutual heavy-light interactions at high density \cite{kondo_d} 
and in strong magnetic fields \cite{kondo_m}. 
Whether and how effects of the enhanced coupling strengths reveal themselves
in the form of $D$-meson properties needs future investigations. 
We plan to consider these topics in the future.

\section*{Acknowledgements}
The authors thank Sungtae Cho and Kenji Morita for useful discussions. 
K.H. thanks Jorge Noronha for a discussion. 
This work was supported by the Korean Research Foundation under
Grants No.~KRF-2011-0020333 and No. KRF-2011-0030621. 
The research of K.H. is supported by JSPS Grants-in-Aid No.~25287066. 
At an early stage of this work, P.G. was supported by the RIKEN Foreign Postdoctoral 
Researcher Program, the RIKEN iTHES Project. 
Four of the authors (P.G., K.H., S.O. and K.S.) thank the Yukawa Institute for Theoretical
Physics, Kyoto University, where some parts of this work were discussed
during the YIPQS international workshops ``{\it New Frontiers in QCD 2013}'' 
and ``{\it Hadrons and Hadron Interactions in QCD 2015}''. 
K.H. is grateful to the hospitality of ECT$^\ast$ where 
this work was finalized during the ECT$^\ast$ workshop 
``New perspectives on Photons and Dileptons in Ultrarelativistic Heavy-Ion Collisions at RHIC and LHC".
This work was partially supported by KAKENHI under Contract No. 25247036.
K.~S. was supported by Grant-in-Aid for JSPS Fellows from Japan Society for
the Promotion of Science (JSPS) (No.26-8288).

\appendix

\section{Estimates of mixing strengths}

\label{sec:g}

\subsection{Mixing strengths extracted from measured radiative decay widths}

\subsubsection{Mixing between charged mesons}

By using the effective Lagrangian of Eq.~(\ref{eq:L_pv}), we obtain an expression for 
the radiative decay width given by \cite{PRD} 
\begin{eqnarray}
\Gamma[ V\rightarrow P + \gamma]
&=& \frac{1}{12} \frac{ e^2 g_{\pv}^{2} \tilde p^{3} }{  \pi m_0^{2} }
\ ,
\end{eqnarray}
where $m_0 = (m_{\ps} + m_{\V})/2$ and $\tilde p = (m_\V^2 - m_\ps^2) / (2 m_\V)$. 
From the measured radiative decay widths, 
we can estimate the mixing strength $g_{\pv}$ between the pseudoscalar and vector states as 
\beq
g_{\pv} &=&  \sqrt{ 12 \pi e^{-2} \tilde p^{ - 3 }  m_0^{2} \; 
\Gamma [V \to P + \gamma  ] \; }
\label{eq:rad}
\ .
\eeq

By inserting the measured decay width for the charged $D$ mesons 
$\Gamma  [ D^{\ast \pm} \to D^\pm + \gamma  ] = 1.536$ keV, we obtain the mixing strength between 
the charged $D^\ast$ and $D$ mesons to be 
\begin{eqnarray}
g_\pv [D^{\ast \pm} \rightarrow D^\pm + \gamma ] 
= 0.9366 .
\label{eq:PVcharged}
\end{eqnarray}

\subsubsection{Mixing between neutral mesons}

For the neutral $D$ mesons, the total decay width has not been measured. 
Nevertheless, the branching ratios in the radiative and hadronic decay modes are known to be 
$ \Gamma[D^{\ast 0} \rightarrow D^0 + \gamma ] : \Gamma[D^{\ast 0} \rightarrow D^0 + \pi^0]
= 38.1 : 61.9 $ \cite{PDG}. 
Therefore, we can obtain the coupling constant involved in the radiative decay mode 
if we know the coupling constant of the hadronic decay $D^{\ast 0} \rightarrow D^0 + \pi^0$. 
From the isospin symmetry, this coupling constant may be related to 
the corresponding hadronic decay modes of the charged $D^{\ast \pm}$: 
\begin{eqnarray}
\label{eqeq}
g[D^{\ast 0} \rightarrow D^0 + \pi^0] &=& g[D^{\ast \pm} \rightarrow D^\pm + \pi^0],
\\
g[D^{\ast 0} \rightarrow D^0 + \pi^0] &=& \frac{1}{\sqrt{2}} g[D^{\ast \pm} \rightarrow D^0 + \pi^\pm]
\ .
\end{eqnarray}

By using the hadronic effective Lagrangian 
\begin{eqnarray}
\Lag &=& g \, \pi  (\partial^\mu D) D^\ast_\mu 
\label{eq:L_DDpi}
,
\end{eqnarray}
we next compute the imaginary part of the $D$--$\pi$ loop, 
and obtain the expression of the hadronic decay width 
\begin{eqnarray}
\Gamma[D^\ast \rightarrow D + \pi] 
&=&  \frac{g^2}{192\pi m_\Dstar^5}  
\{ ( m_\Dstar^2- \mu_+^2 ) ( m_\Dstar^2 - \mu_-^2 ) \} ^{\frac{3}{2}} 
,
\nonumber
\\
\label{eq:G_DDpi}
\end{eqnarray}
where $\mu_\pm = m_D\pm m_\pi$. 
Inserting the measured data $\Gamma[D^{\ast \pm} \rightarrow D^\pm + \pi^0 ] 
=  0.307 \times 83.4 \times 10^{-3} \, {\rm MeV}$, $m_{\pi^0} = 134.9766 \, {\rm MeV}$, 
$m_{D^{\ast \pm}} = 2010.26 \, {\rm MeV} $, and $ m_{D^\pm} = 1869.61\, {\rm MeV}$ 
into Eq.~(\ref{eq:G_DDpi}), 
we find 
\begin{eqnarray}
g[D^{\ast \pm} \rightarrow D^\pm + \pi^0 ] = 11.8592,
\end{eqnarray}
and, by substituting another decay channel $\Gamma[D^{\ast \pm} \rightarrow D^0 + \pi^\pm ] 
=  0.677 \times 83.4 \times 10^{-3} \, {\rm MeV}$, 
and $m_{\pi^\pm} = 139.57018  \, {\rm MeV}$, we obtain  
\begin{eqnarray}
g[D^{\ast \pm} \rightarrow D^0 + \pi^\pm ] = 16.8172.
\end{eqnarray}

Therefore, by utilizing the measured branching ratio mentioned in the paragraph above Eq.~(\ref{eqeq}), 
we obtain the coupling constant of the decay mode 
$ D^{\ast 0} \rightarrow D^0 + \pi^0$ consistently from the first and second decay mode, as    
\begin{eqnarray}
\Gamma[D^{\ast 0} \rightarrow D^0 + \gamma ]  &=& 22.4878 \, {\rm keV},
\\
\Gamma[D^{\ast 0} \rightarrow D^0 + \gamma ]  &=& 22.6108 \, {\rm keV},
\end{eqnarray}
respectively. By using the formula for the radiative decay width, given in Eq.~(\ref{eq:rad}), 
we finally obtain the mixing strength of the neutral $D$ meson as 
\begin{eqnarray}
g [D^{\ast 0} \rightarrow D^0 + \gamma ]  &=& 3.66855, \label{eq:PVfrompions1}
\\
g [D^{\ast 0} \rightarrow D^0 + \gamma ]  &=& 3.67857. 
\label{eq:PVfrompions}
\end{eqnarray}
While the above mixing strengths are obtained on the basis of 
the effective Lagrangian (\ref{eq:L_DDpi}) in the leading derivative expansion, 
the higher-derivative terms could contribute to the coupling among 
$\pi$, $D$, and $D^\ast$. To check the uncertainty involved in 
the estimate of the mixing strength of neutral mesons, 
we examine another method below.

\subsection{Mixing strength from the Bethe-Salpeter amplitude}

The mixing strength between $\eta_c$ and $J/\psi$ was calculated by using 
the Bethe-Salpeter amplitude in the heavy-quark limit in an earlier work by some of the present authors~\cite{PRD}. 
Here, we apply the same method to the estimate of the mixing strength between 
charged $D$ and $D^\ast$ mesons, assuming a constituent quark model picture, in which the $u$ and $d$ quark 
can be considered to be nonrelativistic, and the heavy-quark limit can be applied. 
Below, we follow the steps explained in Appendix B1 of Ref.~\cite{PRD} and generalize them 
to the case of $D$ and $D^\ast$ mesons. 

\begin{figure}[t!]
\begin{center}
\includegraphics[width=0.5\textwidth]{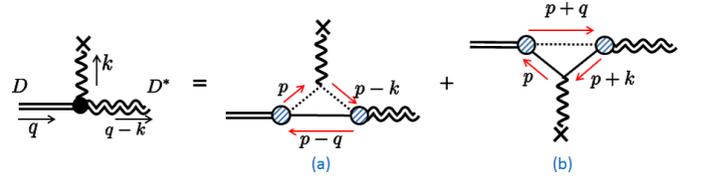}
\end{center}
\vspace{-0.5cm}
\caption{The diagrams needed for computing the mixing strength between 
charged $D$ and $D^\ast$ mesons from the Bethe-Salpeter amplitude.}
\label{fig:triangle}
\end{figure}

To study the mixing strength in this framework, we need to consider the two diagrams depicted in 
Fig.~\ref{fig:triangle}. 
The shaded areas in this figure correspond to form factors given by the 
Bethe-Salpeter amplitudes \cite{YSL}: 
\begin{eqnarray}
\Gamma_{5}(p,p-q) 
&= \Bigl(\epsilon_0 + \frac{\bm{p}^2}{2\mu} \Bigr) \sqrt{\frac{m_h + m_l}{N_c}} 
\Psi_{1S}(\bm{p}) P_{+} \gamma^5 P_{-}, \nonumber \\
& 
\\
\Gamma^{\mu}(p,p-q) 
&= \Bigl(\epsilon_0 + \frac{\bm{p}^2}{2\mu} \Bigr) \sqrt{\frac{m_h + m_l}{N_c}} 
\Psi_{1S}(\bm{p}) P_{+} \gamma^{\mu} P_{-}, \nonumber \\
& 
\end{eqnarray}
where $P_{\pm}$ are the projection operators 
\begin{eqnarray}
P_{\pm} = \frac{1}{2} (1 \pm \gamma^{0}). 
\end{eqnarray}
We take $q = (q_0, 0, 0, 0)$, 
$\Psi_{1S}(\bm{p})$ is the S-wave ground-state wave function, $\epsilon_0$ stands 
for the binding energy of the system ($q_0 = m_l + m_h - \epsilon_0$) and $\mu$ represents the reduced mass of the two quarks: 
$1/\mu = 1/m_h + 1/m_l$. 
Also note that $m_l$ here should be interpreted as the constituent rather than the current 
quark mass. 
The two diagrams of Fig.~\ref{fig:triangle} can be given as 
 \begin{eqnarray}
 i \mathcal{M}^{\mu}_{(a)} =&& - eQ_{l}\int \frac{d^4 p}{(2 \pi)^4} \mathrm{Tr}
\Bigl[
\Gamma^{\mu}(p,p-q) S_{0}(p-q) \nonumber \\ 
&& \times
\Gamma_5^{\dagger}(p-q,p-k) S_{0}(p-k) \gamma_{\nu} 
S_{0}(p) 
\Bigr] A^{\nu}
\end{eqnarray}
and
\begin{eqnarray}
 i \mathcal{M}^{\mu}_{(b)} =&& - eQ_{h} \int \frac{d^4 p}{(2 \pi)^4} \mathrm{Tr} 
 \Bigl[ 
\Gamma_{5}^{\dagger}(p+k,p+q) S_{0}(p+q) \nonumber \\
&& \times \Gamma^{\mu}(p+q,p) S_{0}(p) \gamma_{\nu} 
S_{0}(p+k) 
\Bigr] A^{\nu}. 
\end{eqnarray}
Here, we shall shortly discuss the essential steps for evaluating $i \mathcal{M}^{\mu}_{(a)}$. 
It is convenient to compute first the integral over $p_0$. Completing the contour on either the upper or 
lower part of the complex $p_0$ plane, one picks up two poles. By looking at the respective residues, 
one finds that one pole gives the dominant contribution, while the other is suppressed in the heavy-quark mass limit. 
The dominant pole is found at 
\begin{eqnarray}
p_0 \simeq  m_l - \Bigl(\epsilon_0 + \frac{\bm{p}^2}{2 m_h} \Bigr). 
\end{eqnarray}
Retaining the contribution of this pole while keeping only terms up to first order in $k$, we get 
\begin{eqnarray}
 i \mathcal{M}^{\mu}_{(a)} =&& \frac{m_h + m_l}{m_l} eQ_{l} 
\int \frac{d^3 \bm{p}}{(2 \pi)^3} |\Psi_{1S}(\bm{p})|^2 \epsilon^{0\mu \alpha \nu} k_{\alpha} A_{\nu} \nonumber \\
=&& \frac{m_h + m_l}{m_l} eQ_{l} \tilde{F}^{0\mu} 
\int \frac{d^3 \bm{p}}{(2 \pi)^3} |\Psi_{1S}(\bm{p})|^2 \nonumber \\
=&& \frac{m_h + m_l}{m_l} eQ_{l} \tilde{F}^{0\mu}. 
\end{eqnarray}
The last line is derived from the normalization of the wave function. 
The second diagram of Fig.~\ref{fig:triangle} can be computed in a similar way. The final result reads 
\begin{eqnarray}
 i \mathcal{M}^{\mu}_{(b)} = \frac{m_h + m_l}{m_h} eQ_{h} \tilde{F}^{0\mu}. 
\end{eqnarray}
Adding both terms, we obtain the full amplitude as 
\begin{eqnarray}
 i \mathcal{M}^{\mu} =&&  i \mathcal{M}^{\mu}_{(a)} + i \mathcal{M}^{\mu}_{(b)} \nonumber \\
= && \Bigl( \frac{m_h + m_l}{m_l} Q_{l} +  \frac{m_h + m_l}{m_h} Q_{h} \Bigr) e\tilde{F}^{0\mu}.
\end{eqnarray}
Note that for $m_h = m_l$ this agrees with the results for charmonium of Ref.~\cite{PRD}. 
From the above result, we can read off the mixing strength between pseudoscalar and vector $D$ mesons: 
\begin{eqnarray}
g_{\pv} = \frac{m_h + m_l}{m_l} Q_{l} +  \frac{m_h + m_l}{m_h} Q_{h}.
\label{eq:PV}
\end{eqnarray}
Examining this expression, it becomes clear that $g_{\pv}$ will be dominated by its ligth-quark component, because of the 
large factor $(m_h+m_l)/m_l$. 
Using Eq.~(\ref{eq:PV}), we can compute the ratio of $g_{\pv}$ for neutral and charged D mesons. 
From the above consideration, we expect the absolute value of this ratio to be close to 2, corresponding to the ratio of the absolute charge 
values of $u$ and $d$ quarks. For realistic quark masses, the subleading term can, however, not be ignored. Specifically, we get 
\begin{eqnarray}
\Big| \frac{g_{\pv}(D^{0})}{g_{\pv}(D^{\pm})} \Big| = 3.5, 
\label{eq:PVratio}
\end{eqnarray}
with $m_l=0.3\,\mathrm{GeV}$ and $m_h=1.8\,\mathrm{GeV}$. 

We will now use the above ratio to get an independent estimate for the mixing strength of the neutral D meson, 
which at present cannot be quantified directly as the 
$\Gamma [D^{\ast 0} \rightarrow D^0 + \gamma ]$ decay width has not yet been measured. 
Using Eqs.~(\ref{eq:PVcharged}) and (\ref{eq:PVratio}), we obtain 
\begin{eqnarray}
|g_{\pv}(D^{0})| = 3.278, 
\end{eqnarray}
which agrees quite well with the estimates of Eqs.~(\ref{eq:PVfrompions1}) and (\ref{eq:PVfrompions}), given in the 
previous subsection. 
We will employ these previous estimates in our computations described in the main text of the paper.

\section{Dimension--4 $(eB)^2$ terms}

\label{sec:eB2app}

In this appendix, we briefly summarize the computation of the Wilson coefficients for 
the dimension-4 $(eB)^2$ terms. 
One of the most important observations here is the emergence of the infrared divergences 
when taking the chiral limit for the light quarks. 
This is naturally expected due to the insertions of constant magnetic fields as soft external lines.

To compute the Wilson coefficients corresponding to the graphs shown in Fig.~\ref{fig:eB2}, 
we employ the Fock-Schwinger gauge for the external magnetic field. 
In this gauge, quark propagators with insertions of external magnetic fields are found to be 
\begin{eqnarray}
&& \hspace{-0.6cm}
S^{(0)} (p) = \frac{ i }{ p^2-m^2 } \ ( \slashed p + m ), 
\label{eq:S0}
\\
&& \hspace{-0.6cm}
S^{(1)} (p) = - \frac{ Q_\EM eB }{ (p^2-m^2)^2 } \ ( \slashed p_\parallel + m ) \gtt,
\label{eq:S1}
\\
&& \hspace{-0.6cm}
S^{(2)} (p) = \frac{ 2i Q_\EM^2 (eB)^2 }{ (p^2-m^2)^4 } 
\{ p_\perp^2 (\slashed p + m) -  (p^2-m^2) \slashed p_\perp  \},
\label{eq:S2}
\end{eqnarray}
where $Q_\EM$ denotes the electric charge of a quark in units of ``$e$''. 
We have furthermore 
assumed the magnetic field to be oriented in the third spatial direction. 
The above expressions can be obtained either by inserting external-field lines 
between the free propagator of Eq.~(\ref{eq:S0}) \cite{RRYrev}
or expanding the full propagator given in the proper-time representation with respect to $eB$ 
\cite{Sch, NSVZrev, QED}. 
With Eqs.~(\ref{eq:S0}) - (\ref{eq:S2}), the Wilson coefficients of the dimension-4 $(eB)^2$ terms represented 
in Fig.~\ref{fig:eB2} are given by 
\begin{eqnarray}
&& \hspace{-0.5cm}
\Pi_{\eB}^{(a)} (q) = i N_c \! \int \!\! \frac{d^4p}{(2\pi)^4} 
\tr[ \, \gamma^5 S_l^{(1)} (p) \gamma^5 S_h^{(1)} (q-p) \,], 
\label{eq:eB2a}
\\
&& \hspace{-0.5cm}
\Pi_\eB^{(b)} (q) = i N_c \! \int \!\! \frac{d^4p}{(2\pi)^4} 
\tr[ \, \gamma^5 S_l^{(0)} (p) \gamma^5 S_h^{(2)} (q-p) \,], 
\label{eq:eB2b}
\\
&& \hspace{-0.5cm}
\Pi_\eB^{(c)} (q) = i N_c \! \int \!\! \frac{d^4p}{(2\pi)^4} 
\tr[ \, \gamma^5 S_l^{(2)} (p) \gamma^5 S_h^{(0)} (q-p) \,], 
\label{eq:eB2c}
\end{eqnarray}
where the subscripts $h$ and $l$ on the propagators stand for heavy and light quarks, 
respectively, and the number of the color degrees of freedom $N_c$ comes from 
the trace of the unit color matrix $\tr [\id_c] = N_c$.

As in a standard diagram computation, 
one can perform the loop integrals in Eqs.~(\ref{eq:eB2a})--(\ref{eq:eB2c}) 
with the help of the Feynman parameters. 
Carrying out the momentum integrals, we find 
\begin{widetext}
\begin{eqnarray}
&&
\Pi_\eB^{(a)} =
\frac{  4 N_c } {(4\pi)^2 }  Q_l Q_h (eB)^2 
\int_0^1 \!\! dx \,
x(1-x)  \left[ \,  \Delta^{-1} 
+ ( \, q_\parallel^2 x(1-x) + \ml\mh  \, ) \Delta^{-2} \, \right], 
\\
&&
\Pi_\eB^{(b)}
= 
 \frac{ 4 N_c } { 3 (4\pi)^{2} } Q_h^2 ( e B )^2 \int_0^1 \!\! dx \, \left[\, 
-3 x^2 (1-x) \Delta^{-1} 
+ x^3 \{ \,  q^2 x(1-x) - 2 q_\perp^2 (1-x)  (3-2x) \,\} \Delta^{-2} 
\right.
\nonumber
\\
&&  \hspace{5.2cm} \left.
- 2 q_\perp^2 q^2 x^4 (1-x)^3 \Delta^{-3} 
+ \ml \mh  x^3 \{ \Delta^{-2} - 2 q_\perp^2  (1-x)^2 \Delta^{-3} \}
\, \right],
\label{eq:wk_b}
\\
&&
\Pi_\eB^{(c)} =
 \frac{ 4 N_c } { 3 (4\pi)^{2} } Q_\ell^2 ( e B )^2 \int_0^1 \!\! dx \, \left[ \, 
-3  x (1-x)^2 \Delta^{-1} 
+ (1-x)^3  \{ \,  q^2 (1-x) x - 2 q_\perp^2 x(1 + 2x) \,\} \Delta^{-2} 
\right.
\nonumber
\\
&&  \hspace{5.2cm} \left.
- 2 q_\perp^2 q^2 x^4 (1-x)^3 \Delta^{-3} 
+ \ml \mh  x^3 \{ \Delta^{-2} - 2 q_\perp^2  (1-x)^2 \Delta^{-3} \}
\, \right], 
\label{eq:wk_c}
\end{eqnarray}
\end{widetext}
where we define $\Delta = x \mh^2 + (1-x) \ml^2 - x(1-x)q^2 $.
Equations~(\ref{eq:wk_b}) and (\ref{eq:wk_c}) are related to each other 
through simultaneous interchanges of the charges 
$\Qh \leftrightarrow \Ql$ and masses $\mh \leftrightarrow \ml $, 
which can easily be understood from the diagrammatic representations in Fig.~\ref{fig:eB2}. 
Note, however, that the masses in $\Delta$ are also interchanged. 
This has not been taken into account in the calculation of Eq.~(B7) in Ref.~\cite{Machado}. 
By taking the chiral limit in the above expressions and carrying out the Feynman integrals, 
we obtain the Wilson coefficients (\ref{eq:eB_a0})--(\ref{eq:eB_c2}). 
These results can be compared with the Wilson coefficients for the gluon condensates at finite density 
given in Eqs.~(C.16)--(C.21) of Ref.~\cite{ZHK}. We find that 
the Wilson coefficients agree with each other 
with appropriate replacements of the operators and color matrices:  
\begin{eqnarray}
&&
\tr[\id_c] \frac{\alpha_\EM}{\pi} F^2 \leftrightarrow 
\tr[t^a t^b] \langle \Omega \vert : \frac{\alpha_s}{\pi}  G^2_{ab} : \vert \Omega \rangle, 
\\
&&
-3 \, 
\tr[\id_c] \frac{\alpha_\EM}{\pi}  q^\mu q^\nu
\left( \,  F_{\mu\alpha} F_{\nu}^{\ \alpha} 
- \frac{g_{\mu\nu}}{4}  F_{\alpha\beta} F^{\alpha\beta} \, \right)
\nonumber
\\
&&
\leftrightarrow
\tr[t^a t^b] \left(  q^2 - 4 \frac{(vq)^2}{v^2}  \right)
\nonumber
\\
&&
\hspace{0.7cm} \times
\langle \Omega \vert \! : \!\! \frac{\alpha_s}{\pi} \!\! \left( \, 
\frac{(vG)_{ab}^2}{v^2}-\frac{G_{ab}^2}{4} \, \right) \!\! : \! \vert \Omega \rangle
\nonumber
\end{eqnarray}
where $\tr[t^a t^b] = \delta^{ab}/2$, $G^2_{ab} = G^{a \mu\nu} G^{b}_{\mu\nu}$, and 
$(vG)^2_{ab} = v_\mu G^{a \mu\nu} G^{b}_{\nu\sigma} v^\sigma$. 
The correspondence becomes clearer, if one notes the following decomposition 
of the twist-2 gluon condensate, 
\begin{eqnarray}
&&
\langle \Omega \vert : \! \frac{\alpha_s}{\pi}  \left( \, 
G_{\mu\alpha}^a G_{\nu}^{b \ \alpha} 
- \frac{g_{\mu\nu}}{4}  G_{\alpha\beta}^a G^{b \alpha\beta}
 \, \right) \! : \vert \Omega \rangle
\\
&& \hspace{0.0cm}
= - \frac{1}{3}
\langle \Omega \vert : \! \frac{\alpha_s}{\pi}  \left( \, 
\frac{(vG)_{ab}^2}{v^2}-\frac{G_{ab}^2}{4} \, \right) \! : \vert \Omega \rangle
\left(  g^{\mu\nu} - 4 \frac{v^\mu v^\nu}{v^2}  \right),
\nonumber
\end{eqnarray}
where $v^\mu$ is the flow vector of the medium. 


\begin{figure*}[!tb]
\begin{minipage}[t]{0.48\hsize}
\begin{center}
\includegraphics[width=\textwidth]{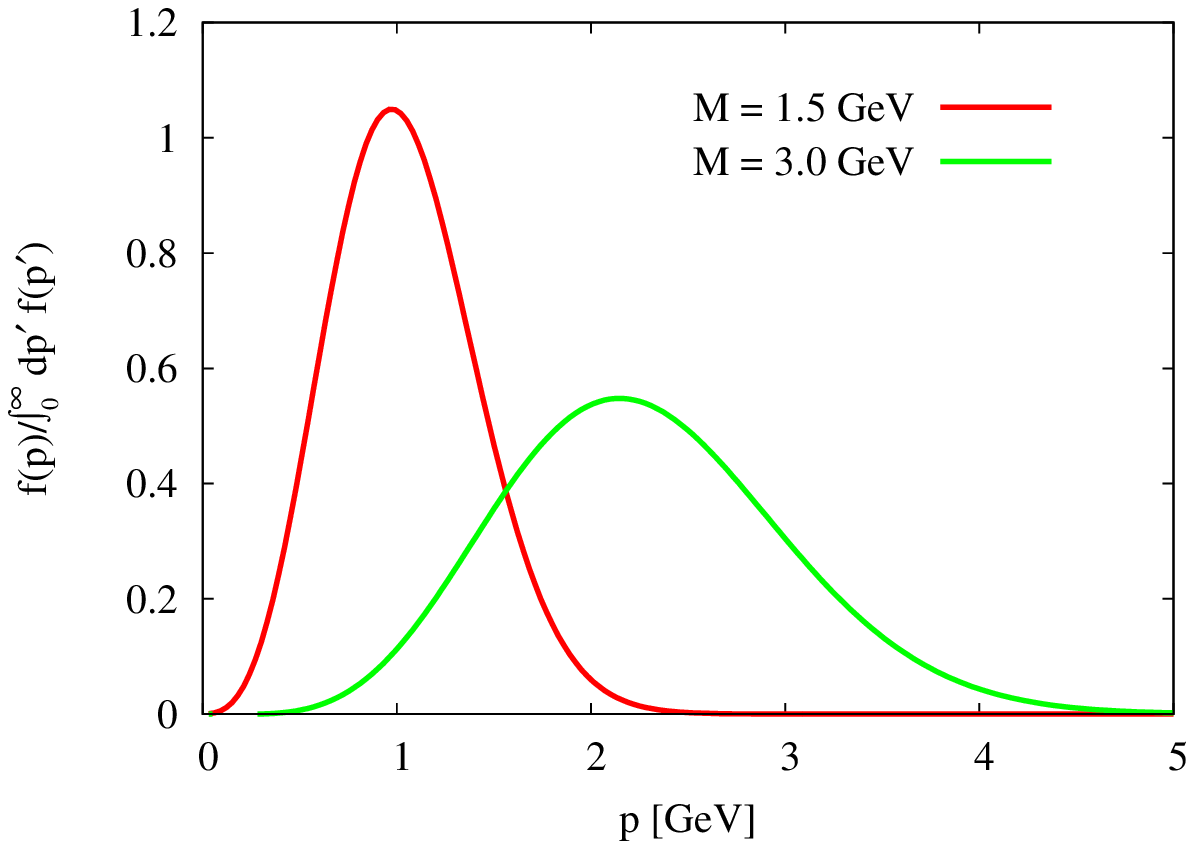}
\end{center}
\vspace{-0.5cm}
\caption{The function $f(p)$ defined in Eq.~(\ref{eq:leading.pert}) for two values of the Borel mass $M$. The curves are 
normalized such that their integrals over $p$ give 1.}
\label{fig:leading.pert}
\end{minipage}
\hspace{0.3cm}
\begin{minipage}[t]{0.48\hsize}
\begin{center}
\includegraphics[width=\textwidth]{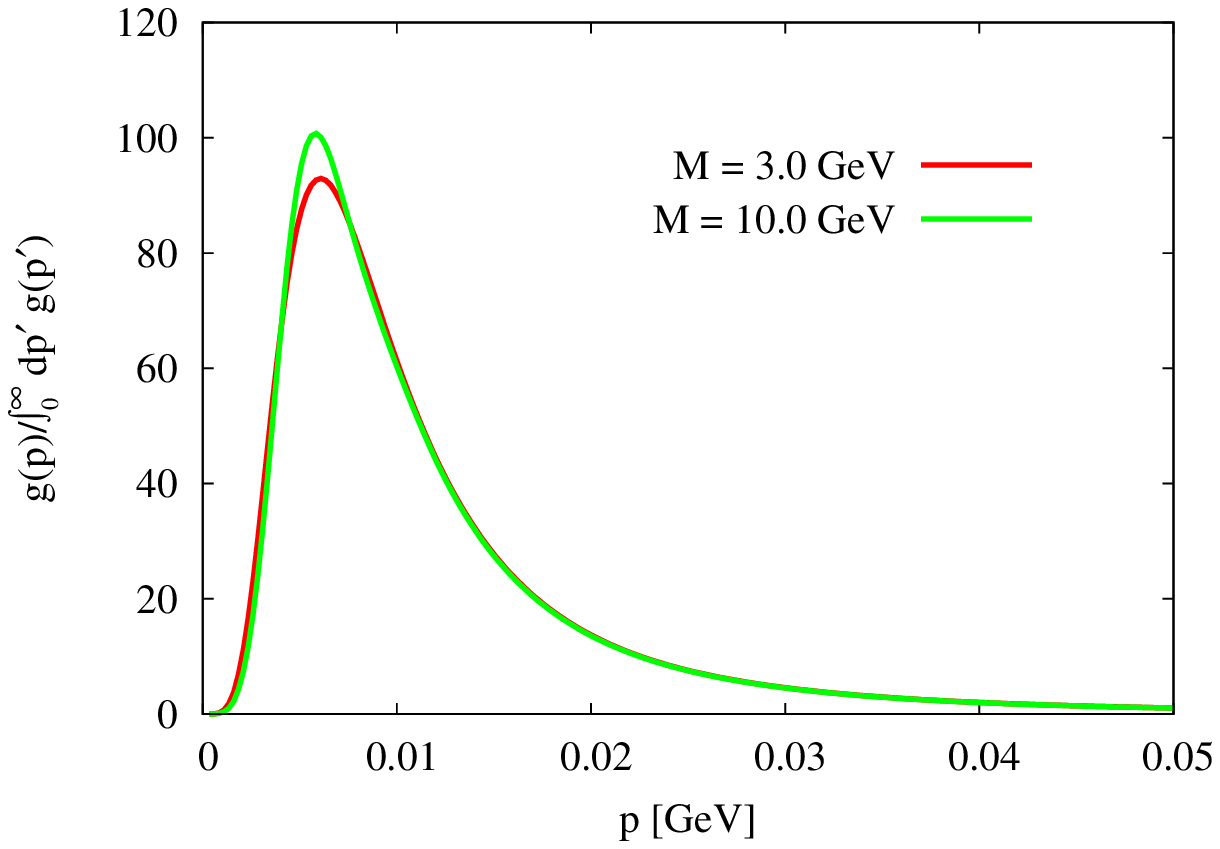}
\end{center}
\vspace{-0.5cm}
\caption{The function $g(p)$ defined in Eq.~(\ref{eq:IR.divergent}) for two values of the Borel mass $M$. The curves are 
normalized such that their integrals over $p$ give 1.}
\label{fig:IR.divergent}
\end{minipage}
\end{figure*}

\section{Nature of IR divergences of perturbative terms}
\label{sec:pertapp}
In this appendix, we will study some of the properties of the IR-divergent terms (terms behaving as $1/m_l$ or 
$\log m_l$ in the chiral limit) and give arguments why they are ``soft" contributions and should 
therefore be considered to be part of the condensates and not the Wilson coefficients. 

For this purpose we will follow the method discussed in Sec. 2.5 of Ref.~\cite{CK2001} for 
studying the properties of the loop integrals. Before considering the IR-divergent terms, let us first 
work out the leading-order perturbative diagram with no external magnetic field attached, which 
will serve as a reference, with which the IR-divergent terms will later be compared. 
It can easily be obtained by replacing, for instance, the first-order propagators $S^{(1)}(p)$ in Eq.~(\ref{eq:eB2a}) 
with the free propagator $S^{(0)}(p)$, given in Eq.~(\ref{eq:S0}). 
After taking the traces and expressing the integrand using a Feynman parameter $x$, we get 
\begin{eqnarray}
4iN_c \int_0^1 dx \int \frac{d^4 p}{(2\pi)^4} \frac{p^2 - x(1-x)q^2 - m_l m_h}{(p^2 - \Delta)^2}. 
\end{eqnarray}
Next, we carry out the Borel transform with respect to $q^2$ before computing the actual loop integral. 
Furthermore, integrating over the angles of the four-dimensional integral and switching to Euclidean 
momenta $p^2$, we obtain, 
\begin{widetext}
\begin{eqnarray}
&&
-\frac{N_c}{2 \pi^2} \frac{1}{M^2} \int_{0}^{\infty} dp \int_0^1 dx \frac{p^3}{x(1-x)} 
\Bigg[ 
1 - \frac{1}{M^2}\Big(\frac{m_l m_h + 2p^2}{x(1-x)} + \frac{m_h^2}{1-x} + \frac{m_l^2}{x} \Big)
\Bigg] e^{-\frac{1}{M^2} \big(\frac{p^2}{x(1-x)}+\frac{m_h^2}{1-x} + \frac{m_l^2}{x} \big)} 
\equiv 
\int_{0}^{\infty} dp f(p)
. 
\nonumber
\\
\label{eq:leading.pert} 
\end{eqnarray}
\end{widetext}
Let us examine the above integrand $f(p)$, 
of which the ultraviolet divergence is regularized by the Borel transform, 
for two different values of the Borel mass $M$. The normalized forms, such that the integrated values 
give 1, are shown in Fig.~\ref{fig:leading.pert}. 
Here and in what follows, we have used $m_h=1.67\,\mathrm{GeV}$ and $m_l = 4.7\,\mathrm{MeV}$ for the quark masses. 
The curves of this figure clearly demonstrate that the loop integral receives its most prominent contribution from momenta 
of the order of $p \sim M$. Therefore, if $M$ is  chosen to be large enough, the momenta running though the 
perturbative loop will be dominated by a ``hard" scale, which allows us to treat them as Wilson coefficients.

Let us now turn to the IR-divergent terms, which, as we will see, have a rather different behavior. To keep the 
discussion short, we will here focus on the term behaving as $1/m_l$ in the chiral limit. 
This term can be extracted from Eq.~(\ref{eq:eB2c}) and corresponds to the second-to-last term in 
Eq.~(\ref{eq:wk_c}). Restoring the momentum integral of this term we have the following expression: 
\begin{eqnarray}
-16iN_c (eB)^2 m_h m_l \int_0^1 dx \int \frac{d^4 p}{(2\pi)^4} \frac{1}{x^5 (1-x)^2} \frac{p^2}{(p^2 - \Delta)^5}. 
\nonumber \\  
\end{eqnarray}
The Borel transform of this term with respect to $q^2$ leads to 
\begin{widetext}
\begin{eqnarray}
\frac{N_c}{12 \pi^2} (eB)^2 \frac{m_h m_l}{M^8} \int_{0}^{\infty} dp \int_0^1 dx \frac{p^5}{x^5 (1-x)^2} 
e^{-\frac{1}{M^2} \big(\frac{p^2}{x(1-x)}+\frac{m_h^2}{1-x} + \frac{m_l^2}{x} \big)} 
\equiv 
\int_{0}^{\infty} dp g(p).
\label{eq:IR.divergent} 
\end{eqnarray}
\end{widetext}
As before, we show the integrand $g(p)$ for two different values of $M$ in Fig.~\ref{eq:IR.divergent}. 
The difference between Figs.~\ref{fig:leading.pert} and \ref{fig:IR.divergent} is apparent. While in Fig.~\ref{fig:leading.pert}, the 
integrand received its largest contribution at a scale of the order of $M$, it is in Fig.~\ref{fig:IR.divergent} determined by 
the light-quark mass and does not much depend on $M$. This shows that the momentum in the loop integral of the IR-divergent term 
is ``soft" and that it should therefore be considered to be part of the condensates and not of the Wilson coefficients.


\section{IR divergences from quark condensates}

\label{sec:condapp}

As discussed in Ref.~\cite{ZHK} and the references cited therein, 
it is more suitable to express the final OPE results in terms of 
quark condensates, which are defined as expectation values of the non-normal-ordered operators instead of the normal-ordered ones. 
The latter one appears naturally in computations based on Wick's theorem and Feynman diagrams. 
Intuitively, the above redefinition can be understood as subtracting the perturbative parts of the condensates, which 
in principle should be included in the Wilson coefficients of the OPE and which can contain IR-divergent terms. 
Specifically, we can relate the 
non-normal-ordered condensates to the normal-ordered ones by the following 
prescription \cite{ZHK,Gro}: 
\begin{eqnarray}
&&
\langle : \bar q \mathcal{O} \Big[ i\vec{D}_{\mu} \Big] q: \rangle = 
\langle \bar q \mathcal{O} \Big[ i\vec{D}_{\mu} \Big] q  \rangle 
\nonumber
\\
&&
+\int \frac{d^4 p}{(2 \pi)^4} \langle : \mathrm{Tr}_{C, D} 
\Bigg( \mathcal{O}\Big[ p_{\mu} +\tilde{A}_{\mu} \Big] S(p) \Bigg):\rangle. 
\label{eq:redef_general}
\end{eqnarray}
Here, $\tilde{A}_{\mu}$ is defined in analogy to Eqs.~(B.11) and (B.12) in 
Ref.~\cite{ZHK} and can at leading order (which is sufficient for 
our purposes) be given as $\tilde{A}_{\mu} \simeq i/2 e F_{\mu \nu} \partial^{\nu}_p$, 
in which the partial derivative is understood to be 
operating with respect to the momentum $p$ and acts only on the propagator $S(p)$. 

Below, we will discuss the redefinition for each of the (light) quark condensates 
that appear in the OPE of this work. 
Linearly and logarithmically divergent terms shown below 
cancel the corresponding divergent terms in the dimension $(eB)^2$ terms 
computed in Appendix~\ref{sec:eB2app}. 
Throughout this appendix, we will use the $\overline{\mathrm{MS}}$ 
scheme to regularize infinite momentum integrals.

\subsection{$\langle \bar q q \rangle$}
For this case, we can simply set $\mathcal{O} = 1$ and hence get 
\begin{eqnarray}
\langle :\bar q q: \rangle = 
\langle \bar q q \rangle + \int \frac{d^4 p}{(2 \pi)^4}  \mathrm{Tr}_{C, D} \big[ S(p)\big].
\end{eqnarray}
For the second term, we next substitute the propagator of Eqs.~(\ref{eq:S0})--(\ref{eq:S2}). 
Taking the trace, it is understood that only 
the leading-order term \big[$S^{(0)}(p)$\big] and the second-order term 
in the magnetic field \big[$S^{(2)}(p)$\big] survive. 
Firstly, we consider the leading-order term, which gives 
\begin{eqnarray}
\int \frac{d^4 p}{(2 \pi)^4} \mathrm{Tr}_{C,D} \big[ S^{(0)}_l(p)\big] 
&=&
 4iN_c m_l \int \frac{d^4 p}{(2 \pi)^4} \frac{1}{p^2 - m_l^2} 
 \nonumber
\\
&=& -\frac{N_c}{4 \pi^2} m_l^3 \left[ \log\Big( \frac{\mu^2}{m_l^2}\Big) + 1\right]. 
 \nonumber
\\
\label{eq:redef_leading}
\end{eqnarray}
As this term is of higher order in $m_l$, it can be ignored in the present discussion. 
Next, we calculate the term of second order in the magnetic field: 
\begin{eqnarray}
&&
\int \frac{d^4 p}{(2 \pi)^4} \mathrm{Tr}_{C,D} \big[ S^{(2)}_l(p)\big] 
\nonumber
\\
&& \hspace{1.4cm}
= 8iN_c m_l Q_l^2 (eB)^2 \int \frac{d^4 p}{(2 \pi)^4} \frac{p_{\perp}^2}{(p^2 - m_l^2)^4} 
\nonumber
\\
&& \hspace{1.4cm}
= \frac{1}{12} \frac{N_c Q_l^2 (eB)^2}{\pi^2} \frac{1}{m_l}. 
\end{eqnarray}
This means that the quark condensate should be redefined as
\begin{eqnarray}
\langle :\bar q q: \rangle = \langle \bar q q  \rangle  
+ \frac{1}{12} \frac{N_c Q_l^2 (eB)^2}{\pi^2} \frac{1}{m_l}. 
\label{eq:redef1}
\end{eqnarray} 
In the above expression, we find the linearly divergent term $1/\ml$.

\subsection{$\langle \bar q \sigma_{12} q \rangle$}
Using again Eq.~(\ref{eq:redef_general}) and setting $\mathcal{O} = \sigma_{12}$, we have 
\begin{eqnarray}
\langle :\bar q \sigma_{12} q: \rangle = 
\langle \bar q \sigma_{12} q  \rangle 
+\int \frac{d^4 p}{(2 \pi)^4}  \mathrm{Tr}_{C, D} \big[\sigma_{12} S(p)\big]. 
\nonumber
\\
\end{eqnarray}
Here, it is noted that, due to the Dirac trace, only the term linear in $eB$ gives a 
nonzero contribution to the second term above. This second term is evaluated as 
\begin{eqnarray}
&&
\int \frac{d^4 p}{(2 \pi)^4} \mathrm{Tr}_{C,D} \big[ S^{(1)}_l(p) \sigma_{12} \big] 
\nonumber
\\
&& \hspace{1.4cm}
= 4iN_c m_l Q_l (eB) \int \frac{d^4 p}{(2 \pi)^4} \frac{1}{(p^2 - m_l^2)^2} 
\nonumber
\\
&& \hspace{1.4cm}
= \frac{1}{4} \frac{N_c Q_l (eB)}{\pi^2} m_l \log \Big( \frac{m_l^2}{\mu^2}\Big), 
\end{eqnarray}
which means that the condensate should be redefined as 
\begin{eqnarray}
\label{eq:redef2}
\langle :\bar q \sigma_{12} q: \rangle = 
\langle \bar q \sigma_{12} q  \rangle 
+  \frac{1}{4} \frac{N_c Q_l (eB)}{\pi^2} m_l \log \Big( \frac{m_l^2}{\mu^2}\Big). 
\nonumber
\\
\end{eqnarray}
It is, however, seen that the second term contains a factor $m_l \log(m_l^2)$, so that it vanishes in the 
chiral limit and we can ignore it in the present calculation. If one needs to consider terms that go beyond the chiral limit, this term 
should of course be taken into account.

\subsection{$\langle \bar q \gamma^{\mu} iD^{\nu} q \rangle$}
Using once more Eq.~(\ref{eq:redef_general}) with $\mathcal{O}$ 
now defined as $\mathcal{O}= \gamma^{\mu} iD^{\nu}$, 
we can rewrite the  condensate as shown below: 
\begin{eqnarray} 
 \label{eq:redef3}
&&
\langle : \bar q \gamma^{\mu} i D^{\nu} q: \rangle 
= \langle \bar q \gamma^{\mu} i D^{\nu} q  \rangle 
\\
&& \hspace{1.2cm}
+ \int \frac{d^4 p}{(2 \pi)^4}  \mathrm{Tr}_{C,D} 
\big[ \gamma^{\mu} (p^{\nu} + \frac{i}{2}e F^{\nu \alpha} \partial_{\alpha} ) S(p) \big]. 
\nonumber
\end{eqnarray} 
We will now consider the second term in more detail. 
The part proportional to $p^{\nu}$ receives nonzero contributions from both $S^{(0)}(p)$ and $S^{(2)}(p)$, 
while the term with $S^{(1)}(p)$ vanishes due to the Dirac trace. Moreover, the contribution involving 
$S^{(0)}(p)$ only leads to a term of order $m_l^4$ similar to Eq.~(\ref{eq:redef_leading}) and can 
therefore be safely neglected. The remaining term gives, 
\begin{widetext}
\begin{eqnarray}
 \int \frac{d^4 p}{(2 \pi)^4}  p^{\nu} \mathrm{Tr}_{C,D} \big[ \gamma^{\mu}  S^{(2)}_l(p) \big]  
&=&
 -\frac{N_c}{3} Q_l^2 (eB)^2 \frac{1}{(4\pi)^2} 
\Bigg[\frac{1}{\epsilon} + \log(4\pi) - \gamma_{\mathrm{E}} - \log ( m_l^2 )  \Bigg] 
 \nonumber
\\
&& \hspace{3.5cm} \times
 \big[ g_{\perp\,\alpha \beta} (g^{\mu \nu}g^{\alpha \beta} 
+  g^{\nu \alpha}g^{\mu \beta} +g^{\nu \beta}g^{\mu \alpha}) -6 g^{\mu\nu}_{\perp}\big] 
\nonumber
\\ 
&=& -\frac{1}{24} Q_l^2 \frac{N_c (eB)^2}{\pi^2} \Bigg[\frac{1}{\epsilon} + \log(4\pi) - \gamma_{\mathrm{E}} - \log (m_l^2)  \Bigg] 
 \big( g^{\mu \nu}_\parallel -   g^{\mu\nu}_{\perp}\big)
, 
\label{eq:redef4} 
\end{eqnarray} 
where we have used $g_{\perp\,\alpha \beta} g^{\alpha \beta} = 2$ in the last line. 
Note that this follows from our implementation 
of the $\overline{\mathrm{MS}}$ scheme, which uses $D_{\parallel}=D-2$ and $D_{\perp}=2$ with $D=4 -2\epsilon$. 
Furthermore, we have for later purposes not carried through the whole regularization procedure at this stage and have hence kept the expression 
with the explicit $1/\epsilon$ pole. 

Next, we need to compute the term proportional to $F^{\nu \alpha}$ in Eq.~(\ref{eq:redef3}). An explicit calculation, 
however, shows that this term exactly vanishes, partly because of the Dirac trace and  partly as a result of the final integration over 
the momentum $p$. We are thus left with Eq.~(\ref{eq:redef4}), which leads to the following redefinition of the condensate, 
\begin{eqnarray} 
\langle : \bar q \gamma^{\mu} iD^{\nu} q: \rangle 
= 
\langle \bar q \gamma^{\mu} iD^{\nu} q \rangle 
- \frac{1}{24} Q_l^2 \frac{N_c (eB)^2}{\pi^2} 
 \big( g^{\mu \nu}_\parallel -   g^{\mu\nu}_{\perp}\big)
\Bigg[\frac{1}{\epsilon} + \log(4\pi) - \gamma_{\mathrm{E}} - \log ( m_l^2 )  \Bigg]. 
\end{eqnarray} 
This is, however, not our final result, as we still need to decompose it into its scalar and symmetric-traceless parts. 
To do this, let us define the second term on the right-hand side of this equation as $C^{\mu \nu}$ and extract from it the relevant Lorentz structures. 
First, the scalar part of $C^{\mu \nu}$ (denoted as $C^{\mu \nu}_{S}$) can be obtained as, 
\begin{eqnarray}
C^{\mu \nu}_{S}=\frac{1}{D} g^{\mu \nu} C^{\alpha}_{\alpha} = \frac{1}{48}  Q_l^2 \frac{N_c (eB)^2}{\pi^2} g^{\mu \nu},
\end{eqnarray}
where we have taken the limit $D \to 4$ after performing the contraction above.
Making use of the Dirac equation, it is easily recognized that this expression is equivalent to Eq.~(\ref{eq:redef1}). 
Next, the symmetric and traceless part of $C^{\mu \nu}$ 
can be obtained by subtracting from it the scalar part 
above as $ C^{\mu \nu}_{ST} = C^{\mu \nu} -C^{\mu \nu}_{S}$. Therefore, we obtain
\begin{eqnarray}
C^{\mu \nu}_{ST} 
&=& - \frac{1}{24} Q_l^2 \frac{N_c (eB)^2}{\pi^2}
 \big( g^{\mu \nu}_\parallel -   g^{\mu\nu}_{\perp}\big)
\Bigg[\frac{1}{\epsilon} + \log(4\pi) - \gamma_{\mathrm{E}} - \log ( m_l^2)  \Bigg] 
- \frac{1}{48}  Q_l^2 \frac{N_c (eB)^2}{\pi^2} g^{\mu \nu}, 
\label{eq:redef3_ST}
\end{eqnarray}
which is regularized to 
\begin{eqnarray}
C^{\mu \nu}_{ST} 
&=&  \frac{1}{24} Q_l^2 \frac{N_c (eB)^2}{\pi^2}
 \big( g^{\mu \nu}_\parallel -   g^{\mu\nu}_{\perp}\big)
\log \Bigl( \frac{m_l^2}{\mu^2}\Bigr) 
- \frac{1}{48}  Q_l^2 \frac{N_c (eB)^2}{\pi^2} g^{\mu \nu}.
\label{eq:redef4_ST}
\end{eqnarray}
In the above expression, we find the logarithmically divergent term $\log \ml^2$. 
\end{widetext}


\begin{figure}[!tbh]
\begin{center}
\includegraphics[clip,width=1.0\columnwidth]{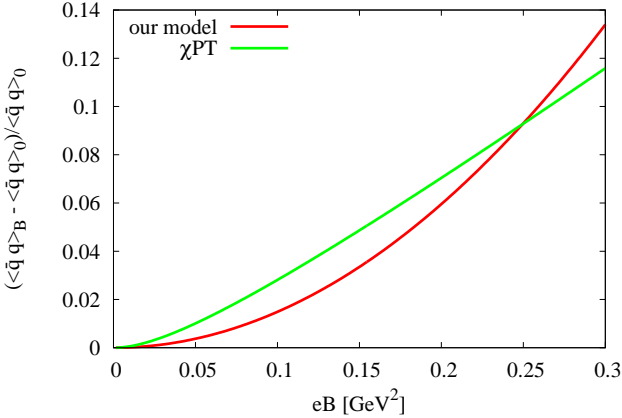}
\caption{The behavior of the chiral condensate $\langle \overline{q} q \rangle_{B}$ as a function 
of the magnetic field $eB$. The red line stands for Eq.~(\ref{eq:quarkcon}), while the green line shows 
the result obtained by chiral perturbation theory~\cite{ChPT2}.}
\label{fig:fig1}
\end{center}
\end{figure}

\section{A simple model for 
the quark condensates in a magnetic field}

\label{sec:simplemodel}

In this appendix, we will discuss a simple model that can qualitatively describe the behavior of the various quark condensates in 
a constant magnetic field. For $\langle \overline{q} q \rangle_{B}$ and $\langle \overline{q} \sigma_{12} q \rangle_{B}$, we 
compare our model results with the findings of chiral perturbation theory and lattice QCD.

\subsection{Method}
We consider a model with quarks that have a constituent quark mass of $m = 300$ MeV. To study the 
modification of any quark condensate, we need to compute the 
traced propagator in a background magnetic field. For treating the possibly divergent 
loop diagrams, we introduce a (Euclidean) cutoff: $\Lambda = 1$ GeV. 
Schematically, our prescription can be summarized as 
\begin{eqnarray}
\hspace{-0.5cm}
\langle \overline{q} \mathcal{O} q \rangle_{B} - \langle \overline{q} \mathcal{O} q \rangle_{0} = -\int^{\Lambda} \frac{d^4 p}{(2\pi)^4} \mathrm{Tr}_{C, D}[\mathcal{O} S(p)_B], 
\end{eqnarray}
where $S(p)_B$ stands for the quark propagator with one or more magnetic field insertions.  Furthermore, 
$\mathcal{O}$ represents a general operator, that can contain gamma matrices or covariant derivatives.

\subsection{$\langle \overline{q} q \rangle_{B}$}
For this most simple case, we can set $\mathcal{O}=1$ and hence get the following expression 
for the leading nonzero term in $eB$, 
\begin{eqnarray}
\langle \overline{q} q \rangle_{B} - \langle \overline{q} q \rangle_{0} 
=
-4i N_c m Q^2 (eB)^2  \!\! \int^{\Lambda} \!\!\! \frac{d^4 p}{(2\pi)^4} \frac{p^2}{(p^2 - m^2)^4}.
\nonumber
\\
\end{eqnarray}
Here, Eqs.~(\ref{eq:S1}) and (\ref{eq:S2}) together with the fact 
that the term linear in $eB$ vanishes were used. 
Evaluating the above integral with the Euclidean cutoff $\Lambda$, the final result reads 
\begin{widetext}
\begin{eqnarray}
\label{eq:quarkcon}
\langle \overline{q} q \rangle_{B} - \langle \overline{q} q \rangle_{0} 
= -\frac{N_c}{12 \pi^2} m Q^2 (eB)^2 \frac{1}{m^2} 
\left[1 - \frac{m^2}{\Lambda^2(1 + m^2/\Lambda^2)^3} \Big(3 + 3\frac{m^2}{\Lambda^2} +  \frac{m^4}{\Lambda^4}\Big) \right]. 
\end{eqnarray}

In Fig.~\ref{fig:fig1}, we compare our result in Eq.~(\ref{eq:quarkcon}) with 
the finding of chiral perturbation theory \cite{ChPT2}, 
for which we have taken the averaged condensate of $u$ and $d$ quarks.  
It is seen in this figure that even though the detailed behavior of our model differs somewhat from that obtained 
with chiral perturbation theory, the overall trend of the two results is in good agreement, as long as one stays 
below $eB=0.3\,\mathrm{GeV}^2$. 

\subsection{$\langle \overline{q} \sigma_{12} q \rangle_{B}$}
Next, we study the condensate $\langle \overline{q} \sigma_{12} q \rangle_{B}$, which vanishes at $eB=0$. 
Setting $\mathcal{O}=\sigma_{12}$, we get at leading order in $eB$, 
\begin{eqnarray}
\langle \overline{q} \sigma_{12} q \rangle_{B} = 
&& 4i N_c m Q (eB) \int^{\Lambda} \frac{d^4 p}{(2\pi)^4} \frac{1}{(p^2 - m^2)^2} 
\nonumber
\\
=&& - \frac{N_c}{4 \pi^2} m Q (eB) \Bigg[
\log\Big( \frac{\Lambda^2}{m^2} \Big) -1 + \log\Big(1+ \frac{m^2}{\Lambda^2} \Big) + \frac{m^2}{\Lambda^2 (1 + m^2/\Lambda^2)}
\Bigg].
\end{eqnarray}
It is noted that the term of $(eB)^2$ in fact vanishes, 
so that corrections to this can at most be of oder $(eB)^3$. 
Translated to the parameter $\tau$ \big[defined as $\langle \overline{q} \sigma_{12} q \rangle_{B} = Q (eB) \tau$\big], the above result gives 
\begin{eqnarray}
\tau =&& - \frac{N_c}{4 \pi^2} m \Bigg[
\log\Big( \frac{\Lambda^2}{m^2} \Big) -1 + \log\Big(1+ \frac{m^2}{\Lambda^2} \Big) + \frac{m^2}{\Lambda^2 (1 + m^2/\Lambda^2)} 
\Bigg] 
\nonumber
\\
=&& -0.0359\,\mathrm{GeV}. 
\end{eqnarray}
Let us now compare this result with the findings of lattice QCD \cite{sus_lat2}: 
\begin{eqnarray}
\tau_u &&= -0.0407 \pm 0.0013\, \mathrm{GeV}, 
\label{eq:tau_u}
\\ 
\tau_d &&= -0.0394 \pm 0.0014\, \mathrm{GeV}. 
\label{eq:tau_d}
\end{eqnarray}
It is seen that our model reproduces the lattice results surprisingly well.

\subsection{$\langle \overline{q} \gamma^{\mu} iD^{\nu} q \rangle_{B}$}
For this condensate, there are no results available 
from either chiral perturbation theory or lattice QCD. 
Therefore, our model will provide a prediction for this case. 
Following the same method as above, we get 
\begin{eqnarray}
\langle \overline{q} \gamma^{\mu} iD^{\nu} q \rangle_{B} =
&&
 -\int^{\Lambda} \frac{d^4 p}{(2\pi)^4} \mathrm{Tr}[\gamma^{\mu}(p^{\nu} + \tilde{A}^{\nu}) S(p)_B] 
\nonumber
\\
=&& \frac{N_c}{24 \pi^2} Q^2 (eB)^2 
 \big( g^{\mu \nu}_\parallel -   g^{\mu\nu}_{\perp}\big)
\Bigg[
\log\Big( \frac{\Lambda^2}{m^2} \Big) -\frac{11}{6} + \log\Big(1+ \frac{m^2}{\Lambda^2} \Big) 
\nonumber
\\
&& \hspace{4.5cm}
+ 3\frac{m^2}{\Lambda^2 (1 + m^2/\Lambda^2)} 
-\frac{3}{2} \frac{m^4}{\Lambda^4 (1 + m^2/\Lambda^2)^2}+\frac{1}{3} \frac{m^6}{\Lambda^6 (1 + m^2/\Lambda^2)^3}
\Bigg] 
\nonumber
\\
&& \hspace{.25cm}
 - \frac{N_c}{8 \pi^2} Q^2 (eB)^2 g^{\mu \nu}_{\perp} \Bigg(
\frac{1}{3}  - \frac{m^2}{\Lambda^2 (1 + m^2/\Lambda^2)} 
+ \frac{m^4}{\Lambda^4 (1 + m^2/\Lambda^2)^2}-\frac{1}{3} \frac{m^6}{\Lambda^6 (1 + m^2/\Lambda^2)^3}
\Bigg)
. 
\end{eqnarray}
This expression is, however, not traceless and therefore contains scalar contributions. We hence have to subtract the trace part. 
Doing this, we obtain the final result:
\begin{eqnarray}
&&
\langle \overline{q} \gamma^{\mu} iD^{\nu} q \rangle_{B} 
=
 \frac{N_c}{24 \pi^2} Q^2 (eB)^2 
 \big( g^{\mu \nu}_\parallel -   g^{\mu\nu}_{\perp}\big) \, A,
\end{eqnarray}
where 
\begin{eqnarray}
A &=& 
\Bigg[
\log\Big( \frac{\Lambda^2}{m^2} \Big) -\frac{4}{3} 
+ \log\Big(1+ \frac{m^2}{\Lambda^2} \Big) 
+ \frac{3}{2}\frac{m^2}{\Lambda^2 (1 + m^2/\Lambda^2)} 
-\frac{1}{6} \frac{m^6}{\Lambda^6 (1 + m^2/\Lambda^2)^3}
\Bigg]. 
\label{eq:A}
\end{eqnarray}



\clearpage

\section{Summary of the Borel transformation}

\label{sec:Borel}

In Tables~\ref{table:OPE_vac}--\ref{table:OPE_c}, 
we summarize the Borel transform of the Wilson coefficients computed in Sec.~\ref{sec:OPE}. 
Useful formulas are available in Ref.~\cite{Coh}. 


\begin{table*}[!h]
\vspace{1cm}
\begin{center}
\begin{tabular}{ l | l }
\hline\hline \\ [-5pt]
LO+NLO & ${
\begin{aligned}
& \mathcal{M}_{\mathrm{pert}}(M^2) = \frac{1}{\pi} \int_{m_h^2}^\infty ds e^{-s/M^2} \mathrm{Im} \, \Pi_{ \mathrm{pert}} (s)  \nonumber\\
& \mathrm{Im} \, \Pi_{ \mathrm{pert}} (s) =  \frac{3}{8\pi} s \left( 1- \frac{m_h^2}{s}\right)^2 \times \left( 1 + \frac{4}{3} \frac{\alpha_s}{\pi} R_0(m_h^2/s) \right) \nonumber\\
& R_0(m_h^2/s = x) = \frac{9}{4} + 2 {\rm Li}_2(x)  +\ln x \ln (1-x) -\frac{3}{2} \ln \frac{1-x}{x} -\ln (1-x) + x \ln \frac{1-x}{x} -\frac{x}{1-x} \ln x 
\end{aligned}}$ \\
\\[-5pt] \hline \\[-5pt]

$\langle \bar{q} q \rangle_\mathrm{vac}$ & $ \mathcal{M}_{\langle \bar{q} q \rangle_\mathrm{vac}}(M^2) = - m_h \langle \bar{q} q \rangle_\mathrm{vac} e^{-m_h^2/M^2} $ \\
\\[-5pt] \hline \\[-5pt]

$ \langle \frac{\alpha}{\pi} G^2 \rangle_\mathrm{vac} $ & $\mathcal{M}_{\langle \frac{\alpha}{\pi} G^2 \rangle_\mathrm{vac}}(M^2) = \frac{ 1}{12} \langle \frac{\alpha}{\pi} G^2 \rangle_\mathrm{vac} e^{-m_h^2/M^2}  $ \\
\\[-5pt] \hline \\[-5pt]

$ \langle \bar{q} g \sigma G q \rangle_\mathrm{vac}$ & $\mathcal{M}_{\langle \bar{q} g \sigma G q \rangle_\mathrm{vac}}(M^2) = \frac{1}{2} \left( \frac{m_h^3}{2M^4} - \frac{m_h}{M^2}\right) \langle \bar{q} g \sigma G q \rangle_\mathrm{vac} e^{-m_h^2/M^2}$ \\
\\[-3pt] \hline \hline
\end{tabular}
\end{center}
\vspace{0.5cm}
\caption{Borel-transformed OPE for $D$ mesons in vacuum \cite{ZHK}. 
The NLO term contains the Spence function ${\rm Li}_2(x) = - \int_0^x \! t^{-1} \ln(1-t) dt $ .}
\label{table:OPE_vac}
\vspace{2cm}
\begin{center}
\begin{tabular}{ l | l }
\hline\hline \\ [-5pt]
$\langle \bar{q} q \rangle_B (eB \leq 0.1 \mathrm{GeV}^2) $ & ${
\begin{aligned}
& \mathcal{M}_{\langle \bar{q} q \rangle_B}(M^2) = -m_h \langle \bar{q} q \rangle_\mathrm{vac} e^{-m_h^2/M^2} \cdot \frac{eB}{16 \pi^2 F_\pi^2} I_H \left( \frac{m_\pi^2}{eB}\right) \\
& I_H(y) = \ln (2\pi) + y \ln \left( \frac{y}{2}\right) - y -2 \ln \Gamma \left( \frac{1+y}{2}\right)
\end{aligned}}$ \\
\\[-5pt] \hline \\[-5pt]
LO$_{(eB)^2}$ & ${
\begin{aligned}
& \mathcal{M}_{(eB)^2}^{(0), D^0}(M^2) = \frac{1}{27} \frac{N_c}{ \pi^2} (eB)^2 e^{-m_h^2/M^2} \\
& \mathcal{M}_{(eB)^2}^{(2), D^0}(M^2) = \frac{3N_c}{4 \pi^2} (eB)^2 \, e^{-m_h^2/M^2}  
\times \frac{4}{81} \left[ \left( 2 + \ln \frac{\mu^2 m_h^2}{M^4} +2 \gamma_E \right) 
\left( 1 -\frac{m_h^2}{M^2} \right) + \frac{2m_h^2}{M^2}   \right]  \\
\end{aligned}}$ \\
\\[-5pt] \hline \\[-5pt]

$\langle \bar{q} \sigma_{12} q \rangle_B$  & $\mathcal{M}_{\langle \bar{q} \sigma_{12} q \rangle}^{D^0}(M^2) = (-\tau_u) (eB)^2 \left( \frac{4}{9} \frac{ m_h }{M^2} - \frac{2}{9} \frac{ m_h^3}{M^4} \right) e^{-m_h^2/M^2} $  \\
\\[-5pt] \hline \\[-5pt]

$\langle \bar{q} \gamma^\mu i D^\nu q \rangle_B$  & $\mathcal{M}_{\langle \bar{q} \gamma^\mu i D^\nu q \rangle}^{D^0}(M^2) = \frac{N_c}{27  \pi^2} (eB)^2 A \left( -1 + \frac{m_h^2}{M^2} \right) e^{-m_h^2/M^2} $ \\

\\[-3pt] \hline \hline
\end{tabular}
\end{center}
\vspace{0.5cm}
\caption{Magnetic-field dependent parts of the Borel-transformed OPE for neutral $D$ mesons. 
$\langle \bar{q} q \rangle_B (eB \leq 0.1 \mathrm{GeV}^2) $ is given by Ref.~\cite{ChPT2}.
Formulas or numerical values for the parameters $\tau_u$ and $A$ are given in Eqs.~(\ref{eq:tau_u}) and (\ref{eq:A}), respectively.}
\label{table:OPE_n}
\end{table*}
\begin{table*}
\begin{center}
\begin{tabular}{ l | l }
\hline\hline \\ [-5pt]
$\langle \bar{q} q \rangle_B (eB \leq 0.1 \mathrm{GeV}^2) $ & 
Common to the OPE for neutral $D$ mesons (see Table~\ref{table:OPE_n}). \\
\\[-5pt] \hline \\[-5pt]

LO$_{(eB)^2}$  & ${
\begin{aligned}
& \mathcal{M}_{(eB)^2}^{(0), D^\pm}(M^2) = -\frac{5}{108} \frac{N_c}{ \pi^2} (eB)^2 e^{-m_h^2/M^2} \\
& \mathcal{M}_{(eB)^2}^{(2), D^\pm}(M^2) = \frac{3 N_c}{4 \pi^2} (eB)^2 
\, e^{-m_h^2/M^2} \left[ \,
\frac{1}{27} \left( 1+\frac{4m_h^2}{M^2} \right) \mathrm{Ei} \left( - \frac{m_h^2}{M^2}\right)
\right. 
\nonumber\\
& \left. \hspace{2.5cm} 
+ \frac{1}{9} + 
\frac{1}{81} \left\{ \left( 2 + \ln \frac{\mu^2 m_h^2}{M^4} +2 \gamma_E \right) 
\left( 1 - \frac{m_h^2}{M^2}  \right) +\frac{2m_h^2}{M^2} \right\}  
\, \right]  
\\
\end{aligned}}$\\

\\[-5pt] \hline \\[-5pt]

$\langle \bar{q} \sigma_{12} q \rangle_B$  & $\mathcal{M}_{\langle \bar{q} \sigma_{12} q \rangle}^{D^\pm}(M^2) = (-\tau_d) (eB)^2 \left( -\frac{2}{9} \frac{ m_h }{M^2} - \frac{1}{18} \frac{ m_h^3}{M^4} \right) e^{-m_h^2/M^2} $ \\

\\[-5pt] \hline \\[-5pt]

$ \langle \bar{q} \gamma^\mu i D^\nu q \rangle_B$   & $
\mathcal{M}_{\langle \bar{q} \gamma^\mu i D^\nu q \rangle}^{D^\pm}(M^2) = \frac{N_c}{108 \pi^2} (eB)^2 A \left( -1 + \frac{m_h^2}{M^2} \right) e^{-m_h^2/M^2} $   \\

\\[-3pt] \hline \hline
\end{tabular}
\end{center}
\vspace{0.5cm}
\caption{Magnetic-field dependent parts of the Borel-transformed OPE for charged $D$ mesons. 
Formulas or numerical values for the parameters $\tau_d$ and $A$ are given in Eqs.~(\ref{eq:tau_d}) and (\ref{eq:A}), respectively.}
\label{table:OPE_c}
\end{table*}

\end{widetext}  



\end{document}